\documentclass[longauth]{aa} 

\usepackage{graphicx}
\usepackage{multirow}
\usepackage{tabularx}
\usepackage{longtable,tabu}
\usepackage{lscape}
\usepackage{subfigure}
\usepackage{natbib}
\usepackage{placeins}
\usepackage[dvipsnames]{xcolor}

\bibpunct{(}{)}{;}{a}{}{,} 
\def\logR{\ensuremath{\log R^{\prime}_{\mathrm{HK}}}}

\begin{document}

\title{The GAPS Programme at TNG}
\subtitle{XXXII. The revealing non-detection of metastable He{\sc i} in the atmosphere of the hot Jupiter WASP-80b\thanks{Based on observations made with the Italian Telescopio Nazionale Galileo (TNG) operated on the island of La Palma by the Fundacion Galileo Galilei of the INAF at the Spanish Observatorio Roque de los Muchachos of the IAC in the frame of the program Global Architecture of the Planetary Systems (GAPS).}} 

\titlerunning{Non-detection of metastable He{\sc i} in the upper atmosphere of WASP-80b}

\authorrunning{Fossati et al.}

\author{L. Fossati\inst{1} \and
        G. Guilluy\inst{2,3} \and
        I. F. Shaikhislamov\inst{4,5,6} \and
        I. Carleo\inst{7,8} \and
        F. Borsa\inst{9} \and
        A. S. Bonomo\inst{2} \and
        P. Giacobbe\inst{2} \and
        M. Rainer\inst{10} \and
        C. Cecchi-Pestellini\inst{11} \and
        M. L. Khodachenko\inst{1,5,12} \and
        M. A. Efimov\inst{4} \and
        M. S. Rumenskikh\inst{4,5,6} \and
        I. B. Miroshnichenko\inst{4,6} \and
        A. G. Berezutsky\inst{4,5} \and
        V. Nascimbeni\inst{8} \and
        M. Brogi\inst{13,14,2} \and
        A. F. Lanza\inst{15} \and
        L. Mancini\inst{16,17,2} \and
        L. Affer\inst{11} \and
        S. Benatti\inst{11} \and
        K. Biazzo\inst{18} \and
        A. Bignamini\inst{19} \and
        D. Carosati\inst{20} \and
        R. Claudi\inst{8} \and
        R. Cosentino\inst{20} \and
        E. Covino\inst{21} \and
        S. Desidera\inst{8} \and
        A. Fiorenzano\inst{20} \and
        A. Harutyunyan\inst{20} \and
        A. Maggio\inst{11} \and
        L. Malavolta\inst{22,8} \and
        J. Maldonado\inst{11} \and
        G. Micela\inst{11} \and
        E. Molinari\inst{23} \and
        I. Pagano\inst{15} \and
        M. Pedani\inst{20} \and
        G. Piotto\inst{22} \and
        E. Poretti\inst{9,20} \and
        G. Scandariato\inst{15} \and
        A. Sozzetti\inst{2} \and
        H. Stoev\inst{20} }

\institute{Space Research Institute, Austrian Academy of Sciences, Schmiedlstrasse 6, 8042 Graz, Austria\\
\email{Luca.Fossati@oeaw.ac.at}
\and
INAF -- Osservatorio Astrofisico di Torino, Via Osservatorio 20, 10025, Pino Torinese, Italy
\and
Observatoire Astronomique de l'Universit\'e de Gen\`eve, Chemin Pegasi 51b, 1290, Versoix, Switzerland
\and
Institute of Laser Physics, SB RAS, Novosibirsk 630090, Russia
\and
Institute of Astronomy, Russian Academy of Sciences, Moscow 119017, Russia
\and
Novosibirsk State Technical University, Novosibirsk 630087, Russia
\and
Astronomy Department, 96 Foss Hill Drive, Van Vleck Observatory 101, Wesleyan University, Middletown, CT, 06459, USA
\and
INAF -- Osservatorio Astronomico di Padova, Vicolo dell'Osservatorio 5, 35122, Padova, Italy
\and
INAF -- Osservatorio Astronomico di Brera, Via E. Bianchi 46, 23807, Merate (LC), Italy
\and
INAF -- Osservatorio Astrofisico di Arcetri, Largo E. Fermi 5, 50125, Firenze, Italy
\and
INAF -- Osservatorio Astronomico di Palermo, P.zza Parlamento 1, I-90134 Palermo, Italy
\and
Lomonosov Moscow State University, Skobeltsyn Institute of Nuclear Physics,119992, Moscow, Russia
\and
Department of Physics, University of Warwick, Gibbet Hill Road, Coventry, CV4 7AL, UK
\and
Centre for Exoplanets and Habitability, University of Warwick, Gibbet Hill Road, Coventry, CV4 7AL, UK
\and
INAF -- Osservatorio Astrofisico di Catania, Via S. Sofia 78, 95123, Catania, Italy
\and
Department of Physics, University of Rome Tor Vergata, Via della Ricerca Scientifica 1, 00133, Roma, Italy
\and
Max Planck Institute for Astronomy, Königstuhl 17, 69117, Heidelberg, Germany
\and
INAF -- Osservatorio Astronomico di Roma, Via Frascati 33, 00040, Monte Porzio Catone (RM), Italy
\and
INAF -- Osservatorio Astronomico di Trieste, via Tiepolo 11, 34143, Trieste, Italy
\and
Fundaci\'on G. Galilei - INAF (Telescopio Nazionale Galileo), Rambla J. A. Fern\`andez P\`erez 7, 38712, Bre\~na Baja (La Palma), Spain
\and
INAF -- Osservatorio Astronomico di Capodimonte, Salita Moiariello 16, 80131, Naples, Italy
\and
Dipartimento di Fisica e Astronomia Galileo Galilei, Universit\`a di Padova, Vicolo dell'Osservatorio 3, 35122, Padova, Italy
\and
INAF -- Osservatorio di Cagliari, via della Scienza 5, 09047, Selargius, CA, Italy
}

\date{Received date ; Accepted date }

\abstract
{Because of its close distance to an active K-type star, the hot Jupiter WASP-80b has been identified as a possible excellent target for detecting and measuring He{\sc i} absorption in the upper atmosphere.}
{Our aim was to look for, and eventually measure and model, metastable He{\sc i} atmospheric absorption.}
{We observed four primary transits of WASP-80b in the optical and near-infrared using the HARPS-N and GIANO-B high-resolution spectrographs attached to the Telescopio Nazionale Galileo telescope, focusing the analysis on the He{\sc i} triplet. We further employed a three-dimensional hydrodynamic aeronomy model to understand the observational results.}
{We did not find any signature of planetary absorption at the position of the He{\sc i} triplet with an upper limit of 0.7\% (i.e. 1.11 planetary radii; 95\% confidence level). We re-estimated the stellar high-energy emission that we combined with a stellar photospheric model to generate the input for the hydrodynamic modelling. We obtained that, assuming a solar He to H abundance ratio, He{\sc i} absorption should have been detected. Considering a stellar wind 25 times weaker than solar, we could reproduce the non-detection only assuming a He to H abundance ratio about 16 times smaller than solar. Instead, considering a stellar wind 10 times stronger than solar, we could reproduce the non-detection only with a He to H abundance ratio about 10 times smaller than solar. We attempted to understand this result by collecting all past He{\sc i} measurements looking for correlations with stellar high-energy emission and planetary gravity, but without finding any.}
{WASP-80b is not the only planet with a sub-solar estimated He to H abundance ratio, suggesting the presence of efficient physical mechanisms (e.g. phase separation, magnetic fields) capable of significantly modifying the He to H content in the upper atmosphere of hot Jupiters. The planetary macroscopic properties and the shape of the stellar spectral energy distribution are not sufficient for predicting the presence or absence of detectable metastable He in a planetary atmosphere, as also the He abundance appears to play a major role.}
\maketitle
\section{Introduction}\label{sec:intro}
Atmospheric escape, that is the process through which planetary atmospheres heat up, expand, and disperse into space, is a fundamental process affecting planetary atmospheric composition, structure, and evolution \citep[e.g.][]{yelle2004,garcia2007,koskinen2010,lopez2013,ildar2014,jin2014,jin2018,owen2017,kubyshkina2018a,modirrousta2020}. For example, it is believed that atmospheric escape has profoundly shaped the evolution of the inner solar system planets and that it has set the basic conditions for the development of a habitable Earth \citep[e.g.][]{lammer2018,lammer2020,airapetian2020}.

Because of the low optical depth of the gas in upper planetary atmospheres, escape is observationally studied typically employing transmission spectroscopy at ultraviolet (UV) wavelengths \citep[e.g.][]{vidal2003,lecavelier2012,fossati2010,linsky2010,ehrenreich2015,bourrier2018c,garcia2021}. However, most planet-hosting stars are dim in the UV and the background stellar light is spatially and temporally variable, particularly in the far-UV \citep[FUV; e.g.][]{haswell2012,llama2015,llama2016}. This often poses challenges to the interpretation of the observations, which have indeed led to controversial results \citep[e.g.][]{vidal2003,videl2008,benjaffel2007,benjaffel2008,benjaffel2010,linsky2010,ballester2015}.

\citet{seager2000} and \citet{oklopcic2018} suggested that the metastable He{\sc i} 2$^3$S triplet at $\approx$10830\,\AA\ may be an alternative way to UV observations for probing upper atmospheres and escape. These features have the significant advantage of lying in a region of the near-infrared (nIR) relatively devoid of absorption lines from the Earth’s atmosphere and close to the peak of the spectral energy distribution of typical planet hosts. Early attempts at detecting these features at low spectral resolution from the ground were unsuccessful \citep{moutou2003}, whereas absorption was detected for the warm giant WASP-107b at low resolution with HST \citep{spake2018}. The He{\sc i} triplet has then been detected for a number of close-in gas giant planets, mostly employing ground-based high-resolution transmission spectroscopy \citep[e.g.][]{nortmann2018,allart2018,allart2019,salz2018,mansfield2018,alonso2019,guilluy2020,paragas2021}.

The observational results of \citet{nortmann2018} indicate that the presence and strength of the He{\sc i} triplet in planetary transmission spectra depend strongly on stellar activity, and in particular on the high-energy stellar radiation (X-ray and EUV; together XUV). \citet{oklopcic2019} studied the formation of the triplet as a function of stellar spectral energy distribution (SED) finding that the formation and strength of the triplet does not depend exclusively on the XUV stellar emission, but also on the near-UV (NUV) emission (i.e. $\lesssim$2600\,\AA), which is the radiation ionising He{\sc i} in the metastable state \citep[see also][]{lampon2020,lampon2021,ildar2021,khodachenko2021a}. Therefore, the He{\sc i} metastable lines are preferentially formed for planets with extended atmospheres orbiting stars with strong XUV and low NUV emission, that is active K-type stars. The non-detection of the He{\sc i} triplet for GJ436b \citep{nortmann2018}, although the atmosphere is heavily escaping \citep[e.g.][]{ehrenreich2015}, demonstrates the importance of the shape of the stellar SED in the formation of these features. 

\citet{allart2019} applied to WASP-107b a three-dimensional (3D) Monte Carlo code to show that, through radiation pressure, the stellar nIR emission plays a significant role in shaping the He{\sc i} planetary absorption. For this planet, \citet{khodachenko2021a} and \citet{wang2021} derived a nearly solar Helium abundance (He/H\,$\approx$\,0.1), although the latter work disregarded radiation pressure, which led them to consider an extreme stellar wind of ten times solar to explain the observed $\approx$3\,km\,s$^{-1}$ blue shift of the planetary absorption features. Instead, \citet{khodachenko2021a} were able to reproduce the observations considering a more moderate solar-like wind thanks to the self-consistent inclusion of the stellar radiation pressure acting on the metastable helium.

The detection and measurement of metastable He{\sc i} gives the unique opportunity to constrain the atmospheric He abundance of exoplanets. Employing a one-dimensional (1D) aeronomy code based on the models developed by \citet{salz2016} and \citet{oklopcic2018}, \citet{ninan2020} and \citet{palle2020} derived for the warm Neptune-like planet GJ3470b a He abundance 5--10 times smaller than solar (i.e. He/H\,$\approx$\,0.01). This low He abundance has been further confirmed by 3D multi-fluid self-consistent aeronomy simulations by \citet[][He/H\,$\approx$\,0.013]{ildar2021}, who further strengthened the importance of the stellar wind for modelling the observed He{\sc i} transit absorption features. Using a 1D hydrodynamic (HD) code, \citet{alonso2019} and \citet{lampon2020} derived a similarly low He abundance also for the hot Jupiter HD209458b. \citet{lampon2021} applied a 1D Parker-like solution fitted to the aeronomy simulation of \citet{salz2016} to model the atmospheric outflow of HD189733b finally obtaining a rather low He abundance of He/H\,$\approx$\,0.008 (i.e. about ten times sub-solar). These results indicate that a non-solar He/H abundance ratio may be a common characteristic among hot Jupiters.

WASP-80b is a hot Jupiter orbiting a K-type star \citep{triaud2013,triaud2015,mancini2014,bonomo2017}, which is rather active \citep[e.g.][]{salz2016,king2018}. Transmission spectroscopy of WASP-80b has been carried out in the optical from the ground and in the nIR with HST. The observations led to a tentative detection of the Na and K alkali lines \citep{sedaghati2017} and of water \citep{tsiaras2018,fisher2018}, suggestive of low metallicity. \citet{salz2016} computed 1D hydrodynamic simulations of the upper atmosphere of WASP-80b concluding that this is one of the most promising targets for the observational detection of atmospheric escape, particularly at FUV wavelengths. More recent estimates of the stellar high-energy emission by \citet{king2018} further strengthened the conclusion of \citet{salz2016}. Therefore, given that the planet orbits an active K-type star and that it is believed to host an extended atmosphere, WASP-80b appears to be an ideal candidate for the search and detection of metastable He{\sc i} from the ground employing high-resolution transmission spectroscopy \citep{kirk2020}.

We present here nIR transmission spectroscopy observations of WASP-80b, carried out with the GIANO-B high-resolution spectrograph \citep{Claudi2017}, covering the He{\sc i} triplet. We also present 3D modelling of the planetary atmosphere attempting to reproduce the observational results and finally aiming at constraining key parameters characterising the planetary upper atmosphere.

This paper is organised as follows. Section~\ref{sec:observations} presents the observations and the data analysis, while in Section~\ref{sec:results} we describe the results obtained from the observations. Section~\ref{sec:model} shows the results obtained from 3D modelling of the upper atmosphere of WASP-80b aiming at reproducing the observations. In Section~\ref{sec:dicsconc} we summarise the work, discuss the observational and theoretical results, putting them in a wider context, and gather the conclusions.
\section{Observations and data analysis}\label{sec:observations}
\subsection{Data Reduction}\label{reduction}
We observed the WASP-80 system with the nIR echelle spectrograph GIANO-B installed on the 3.6~m Telescopio Nazionale Galileo (TNG) telescope \citep{Oliva2006}. The observations were carried out in GIARPS (GIANO-B + HARPS-N; \citealt{Claudi2017}) observing mode and were performed with the nodding acquisition mode, with the target observed at predefined A and B positions on the slit, following an ABAB pattern \citep{Claudi2017}. Therefore, the target and sky spectra are taken in pairs by using the two nodding positions along the slit (A and B); in this way the slit looking at the sky provides an accurate reference for subtracting the thermal background and telluric emission lines.

\begin{figure*}
		\centering
		\includegraphics[width=17cm]{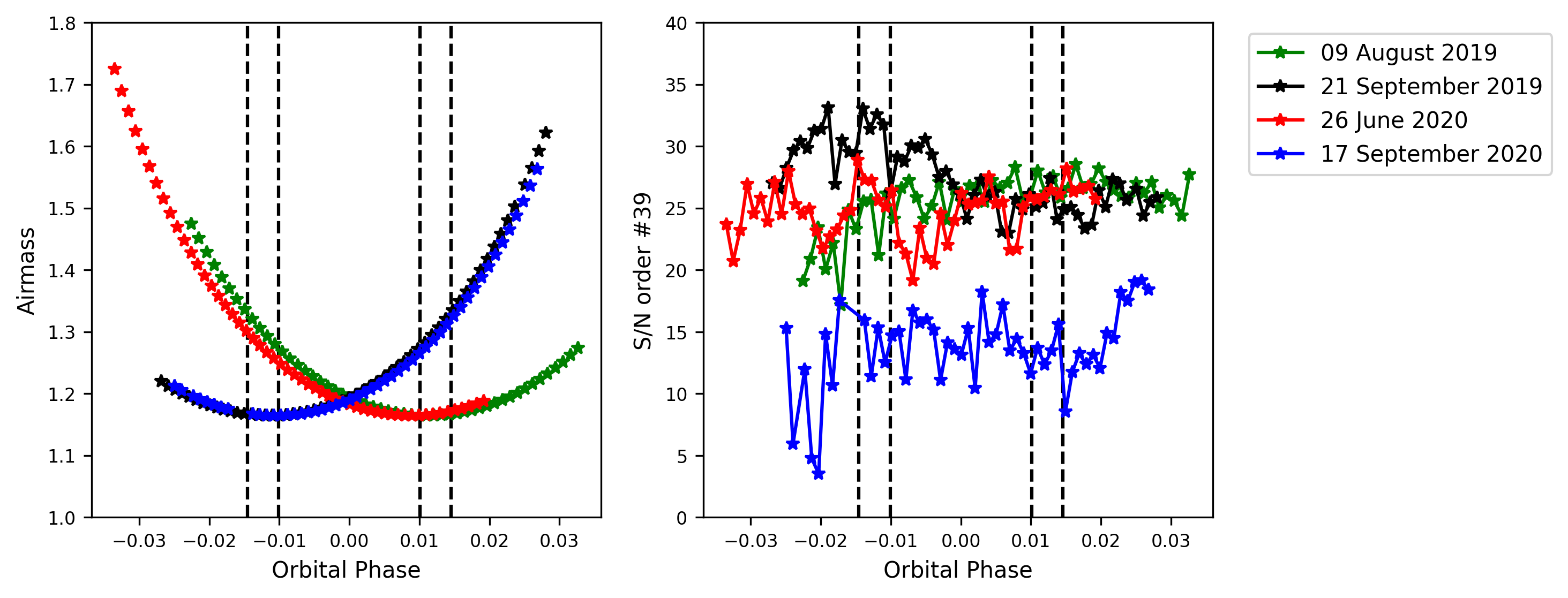}
		\caption{Airmass (left panel) and S/N (right panel) measured during the GIARPS observations. The vertical dashed lines mark the $t_1$, $t_2$, $t_3$, and $t_4$ contact points (from left to right).}
		\label{SNR_}
\end{figure*}

GIANO-B achieves simultaneous coverage in the wavelength range 0.95–2.45\,$\mu$m, split into fifty orders, at a spectral resolving power of $R$\,$\sim$\,50,000. The dataset encompasses four primary transit events (UT 09 August 2019, UT 21 September 2019, UT 26 June 2020, UT 17 September 2020) of WASP-80b that have been observed within the context of the Global Architecture of Planetary Systems (GAPS) programme \citep{borsa2019,guilluy2020,Giacobbe2021}. Table~\ref{tab_log} presents the observing log by listing the number of collected spectra, exposure times, and achieved signal-to-noise ratio (S/N) in the spectral region of interest (10825--10845\,\AA), while Figure~\ref{SNR_} shows the variation of the S/N for each image (for the spectral order 39). As Table~\ref{tab_log} and Figure~\ref{SNR_} show, the data collected during the observed transit event at UT 17 September 2020 exhibit a lower S/N compared to the other observations, probably due to the presence of thin clouds. We thus preferred to discard this transit from the analysis. The target was observed within an airmass range of 1.16--1.72 (see the left panel of Figure~\ref{SNR_}).
\begin{table}
	\caption{WASP-80\,b log of TNG-GIARPS observations.}
	\label{tab_log} 
	\small
	\centering         
	\begin{tabular}{c | c | c | c }          
		\hline\hline                       
		Night &   N$_{\rm obs}$ & Exposure time & S/N$_{\rm{AVE}}$ \\	
		\hline	
		09 August 2019 &  52 &  200s &  28 \\
		21 September 2019 &  56 &  200s &  26 \\
		26 June 2020 &  54 &  200s &  26 \\
		17 September 2020 &  50 &  200s &  18 \\
		\hline 
	\end{tabular}
	\tablefoot{
		The second column gives the number of spectra collected during each night. The fourth column lists the time-averaged S/N in the spectral region containing the He{\sc i} triplet (10825--10845\,\AA).
	}
\end{table}

The GIANO-B spectra were dark-subtracted, flat-field corrected, and extracted (without applying the blaze function correction) with the GOFIO data reduction pipeline \citep{Rainer2018}, which provides also a preliminary wavelength calibration (defined in vacuum) using an U-Ne lamp spectrum as a template. The reduction process includes also bad pixels removal\footnote{We updated the GOFIO bad-pixel mask to account for several bad pixels, which contaminated the spectral region of interest.}. The resulting spectra are in the terrestrial rest frame. For the rest of the analysis we focused on the spectral order 39, where the helium triplet falls.

One spectrum (2459027.627261~BJD$_\mathrm{TDB}$, at phase 0.00475), collected during night UT 26 June 2020, exhibited an anomalous flux excess at $\lambda\sim$10832.74\,\AA\ and another at $\lambda\sim$10832.89\,\AA\ (pixel \#688 and \#689) probably due to a strong cosmic-ray not perfectly corrected by the GOFIO pipeline, thus we preferred to discard this particular spectrum from the rest of the analysis.

The output of the GOFIO pipeline (wavelength solution and flux) required additional processing steps before proceeding with the data analysis. The mechanical instability of the instrument causes the wavelength solution to change during the observations and, since the U-Ne lamp spectrum is only acquired at the end of the observations, the wavelength solution determined by GOFIO could be not particularly accurate. We corrected for this instability by following the recipe described by \citet{Brogi2018}, \citet{Guilluy2019, guilluy2020}, and \citet{Giacobbe2021}. In short, for each observation, we aligned every spectrum to a common wavelength scale using spline interpolation based on the measured shift computed via cross-correlation with a time-averaged observed spectrum of the target used as a template. This means aligning the sequence to the reference frame of the Earth’s atmosphere, which is also assumed as the frame of the observer neglecting any $\sim$10\,m\,s$^{-1}$ differences due to winds. We successively used the atmospheric transmission spectrum generated via the ESO Sky Model Calculator\footnote{\url{https://www.eso.org/observing/etc/bin/gen/form?INS.MODE=swspectr+INS.NAME=SKYCALC}} to refine the standard GOFIO wavelength calibration.
\subsection{Transmission spectroscopy}
%
\begin{table}
		\caption{Stellar and planetary parameters adopted in this work.}             
		\label{tab_par}  
		\footnotesize
		\centering  
		\begin{tabular}{l c l}          
			\hline\hline                       
			Parameters &  Value & Reference \\ 
			\hline 
			\multicolumn{2}{l}{Planetary and transit parameters} & \\
			T$_{\rm 0}$ [BJD$_\mathrm{TDB}$] &2456125.417574(86) &\citet{bonomo2017} \\
			P [d] &  3.06785234$^{+0.00000083}_{-0.00000079}$ & \citet{triaud2015}\\
			i [deg] &  89.02$^{+0.11}_{-0.10}$ & \citet{triaud2015}\\
			b &   0.215$^{+0.020}_{-0.022}$ & \citet{triaud2015} \\
			R$_{\rm P} $ [R$_{\rm {Jup}}$]  & 0.9990$^{+0.0300}_{-0.0310}$ & \citet{triaud2015}\\
			M$_{\rm P} $ [M$_{\rm {Jup}}$]  & 0.538$^{+0.035}_{-0.036}$ & \citet{triaud2015}\\
			$\rho_{\rm P} $ [g\,cm$^{-3}$]  & 0.717$^{+0.039}_{-0.032}$ & \citet{triaud2015}\\
			a [au] & 0.0344$^{+0.0010}_{-0.0011}$  & \citet{triaud2015}\\
			k$_{\rm P}$ [km s$^{-1}$]& 122.0$^{+3.5}_{-3.9}$ & This paper \tablefootmark{(a)}\\
			e & $<$0.02 &  \citet{bonomo2017} \\             
			\hline
			\multicolumn{2}{l}{Stellar parameters} & \\
			R$_\star$  [R$_{\rm {\odot}}$] & 0.586$^{+0.017}_{-0.018}$ & \citet{triaud2015}\\
			M$_\star$  [M$_{\rm {\odot}}$] & 0.577$^{+0.051}_{-0.054}$ & \citet{triaud2015}\\
			k$_{\rm s}$ [m s$^{-1}$] &  109.0$^{+3.1}_{-4.4}$  & \citet{triaud2015}\\
			V$_{\rm sys}$ [km s$^{-1}$] & 9.82(77)  km s$^{-1}$ & \textit{Gaia} DR2\tablefootmark{(b)}\\
		 	$B-V$ &  0.929 & \citet{triaud2013}\\
			\hline                                       
		\end{tabular}
		\tablefoot{From top to bottom, the parameters are time of central transit, planetary orbital period, inclination angle, transit impact parameter, planetary radius, planetary mass, planetary bulk density, orbital separation, semi-major amplitude of the planetary radial velocity curve, eccentricity, stellar radius, stellar mass, semi-major amplitude of the stellar radial velocity curve, systemic radial velocity, Johnson $B-V$ color.
		\tablefoottext{a}{Derived from $a$, $P$, and $i$ as $\frac{2 \pi a}{P}\sin{i}$.}
		\tablefoottext{b}{\citet{gaiadr2}.}
	}
	\end{table}	
We performed transmission spectroscopy applying the steps described below independently to each transit and considering the system parameters listed in Table~\ref{tab_par}. First, we corrected for contamination from the Earth’s atmosphere, which produces both absorption and emission lines in the spectral region around the helium triplet. We corrected for the Earth's absorption lines by using the relation between airmass and strength of the telluric lines \citep[e.g.][]{Snellen2008, Vidal-Madjar2010, Astudillo-Defru2013}. To this end, we first shifted (via quadratic interpolation) each spectrum to the stellar rest frame by computing the stellar radial velocity V$_\star$ in the telluric reference system. Assuming a circular orbit (see Table~\ref{tab_par}), this is given by: 
\begin{equation}
    V_\star=V_{\mathrm{sys}}+V_{\mathrm{bar}}-k_\star \sin[2\pi\,\phi(t)]\,,
\end{equation}
where we accounted for the velocity of the observer induced by the rotation of the Earth and by the motion of the Earth around the Sun (i.e. the barycentric Earth radial velocity, $V_{\mathrm{bar}}$), the stellar reflex motion induced by the planet (i.e. $k_\star \sin[2\pi\,(\phi(t))]$, where $\phi$ is the planet's orbital phase and $k_\star$ is the stellar radial-velocity semi-amplitude), and the systemic velocity of the star-planet system with respect to the barycentre of the solar system ($V_{\mathrm{sys}}$). Then, we normalised each spectrum by dividing it by the average flux within two intervals on the immediate blue (10826.0--10828.0\,\AA) and red (10838.5--10839.5\,\AA) sides of the He{\sc i} triplet, where telluric and stellar lines are absent. 

\begin{figure}
    \centering
    \includegraphics[width=9cm]{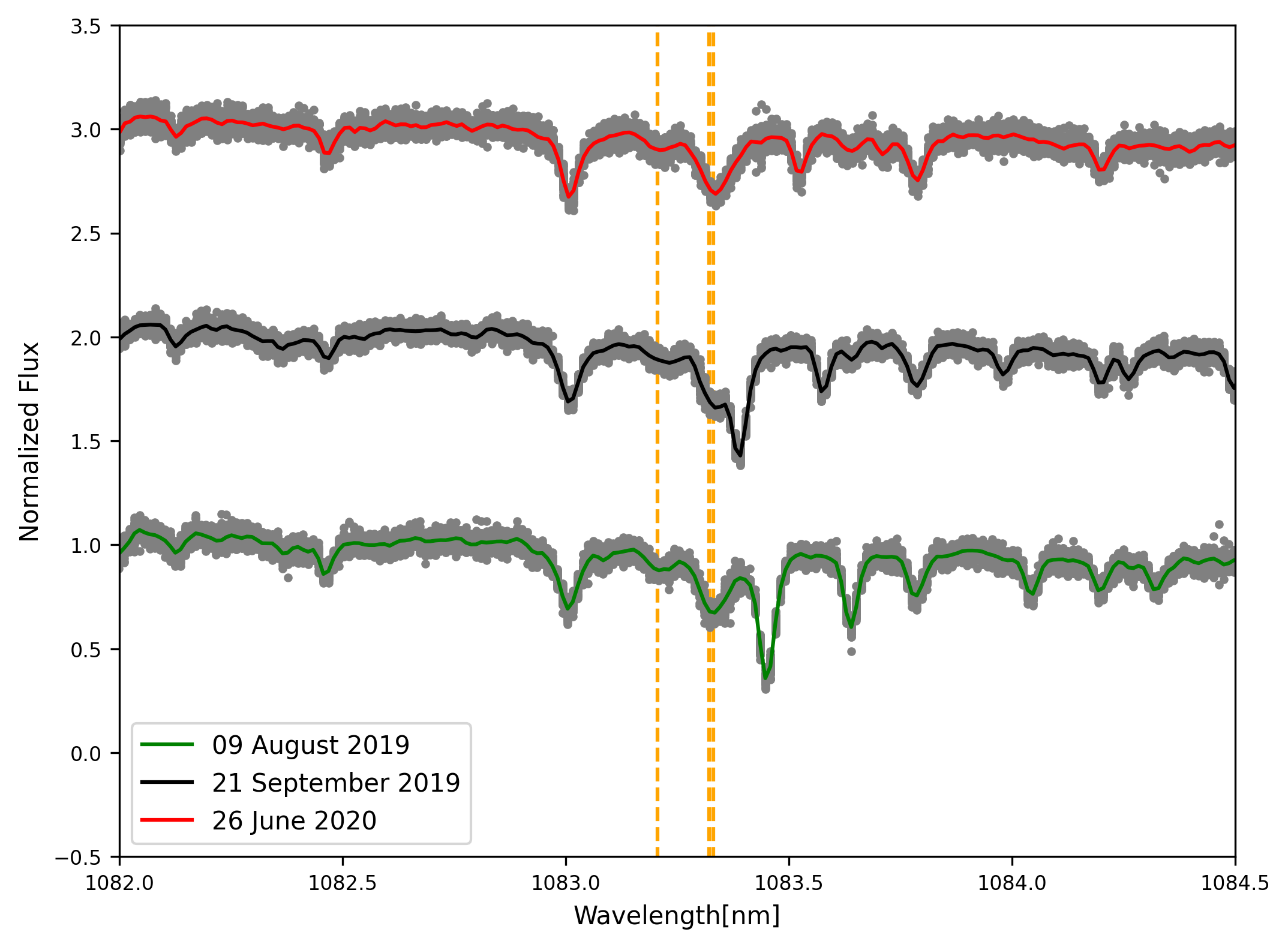}
    \caption{Normalised spectra in the stellar rest frame of the three considered transits (gray dots) overlaid the correspondent time-averaged spectrum. The spectra are plotted with vertical offsets for clarity. Vertical orange lines mark the position of the three components of the metastable He{\sc i} triplet.}
    \label{sp_norm}
\end{figure}

Using the out-of-transit spectra alone, we created a telluric reference spectrum (T($\lambda$)) by extracting the linear correlation existing between the logarithm of the normalised flux and the airmass \citep[e.g.][]{Wyttenbach2015}. Then, we divided all spectra by the reference telluric spectrum rescaled such that each spectrum would have been acquired at the same airmass, namely the average airmass of the in-transit spectra (collected between the t$_1$ and t$_4$ contact points). In this way, we did not directly correct for telluric lines, but bring them to the same strength across all spectra so that they were automatically deleted when we created the transmission spectra. Since the shift of the telluric lines in the stellar rest frame due to the barycentric component in each analyzed transit was much lower than the instrumental resolution, we preferred to perform the telluric removal in the stellar rest frame and not in the Earth's one. In this way, we avoided spurious features in the telluric reference spectrum in correspondence of strong stellar lines because of the low S/N  (see \citealt{Borsa2018}). We remark that the final results do not depend on the rest frame of the telluric correction. Since during the transit at UT 26 June 2020 the telluric contamination was practically absent, we decided to not perform the telluric correction, in this way we avoided correlated noise in the final spectra. Figure~\ref{sp_norm} shows the results of these preliminary reduction steps for each observing night. On the night of the UT 21 September 2019, the strongest component of the helium triplet is blended with the water telluric absorption line at 18835.1\,\AA\ (wavelength in the Earth's rest frame). However, as  the result we obtained from this observing night is in agreement with that obtained in the other two nights, we are confident that the applied telluric removal worked appropriately.

Ground-based observations are contaminated also by telluric emission lines. In particular, in the spectral region of interest, there are three OH emission lines that fall near the He{\sc i} triplet (at $\sim$10832.1\,\AA, $\sim$10832.4\,\AA, and $\sim$10834.3\,\AA, where the wavelengths are in vacuum). However, since the observations have been gathered with the nodding acquisition mode that allows for subtraction of the thermal background and emission lines (see Sect.~\ref{reduction}), there was no need to perform an additional correction, as instead has been done in other works \citep[e.g.][]{nortmann2018,salz2018,allart2019}.

\begin{figure*}
    \centering
    \includegraphics[width=17cm]{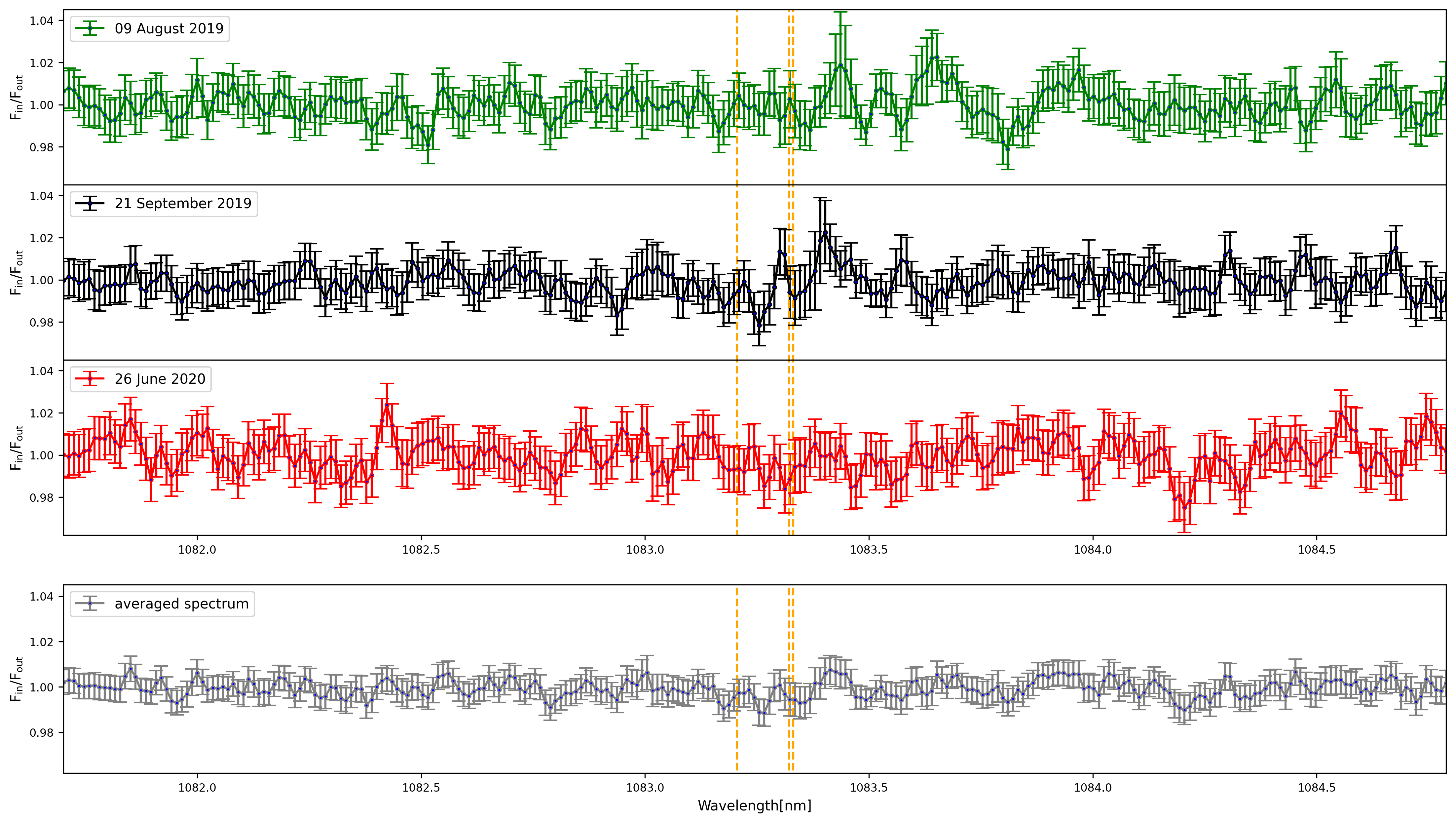}
    \caption{Top: weighted mean averaged transmission spectrum for each observed transit. Bottom: weighted mean of the three observed transits. Vertical orange lines mark the position of the three components of the metastable He{\sc i} triplet. Since the final error bars are calculated following error propagation by taking as initial errors the square root of the extracted spectra aligned in the telluric rest frame, the final error bars correspond to one standard deviation.}
    \label{sp_transmission}
\end{figure*}

For each night, we then built a master stellar spectrum $S_\mathrm{master}$ from all out-of-transit spectra (i.e. with orbital phase smaller than t$_1$ or greater than t$_4$) by computing the weighted mean, using $w$\,=\,1/$\sigma^2$ as the respective weights, where $\sigma$ are the uncertainties associated with each wavelength bin. We then derived the transmission spectra $T$ by dividing each spectrum for $S_\mathrm{master}$. Transmission spectra corrected for the telluric lines with the airmass relation can still present some correlations and telluric residuals caused by the variation of precipitable water vapor. We thus applied a second telluric correction following the approach of \citet{Wyttenbach2015}. In short, for each observation, we performed a linear fit between the telluric reference spectrum (previously calculated, i.e. T($\lambda$)) and the transmission spectrum. We then divided the transmission spectrum by the fit solution. All full in-transit transmission spectra (i.e. obtained between the t$_2$ and t$_3$ contact points) were finally averaged to create the transmission spectrum for each observed transit. The top panel of Figure~\ref{sp_transmission} shows the weighted mean averaged transmission spectrum for each night, while the bottom panel displays the averaged transmission spectrum over the three observed transits $T_\mathrm{ave}$.
We did not correct for the fringing pattern typical of GIANO-B spectra as the modulation caused by it is significantly smaller than the final error bars\footnote{We estimated the amplitude of any possible fringing pattern by fitting sinusoids to the transmission spectra, and verified their amplitude to be less than 29$\%$ of the average error-bar.}. To remove possible linear trends in the continuum, we computed a robust linear fit of each $T$ in the 10815--10850\,\AA\ range and divide it out. We avoided the region around the helium triplet ($\pm 20$\,km\,s$^{-1}$ centered at 1083.3\,nm) in performing the linear fit. Then, we shifted via quadratic interpolation every $T$ in the planetary reference frame, by calculating the planet's radial velocity in the stellar rest frame as
\begin{equation}
    V_{\mathrm{P}}=+k_\star \sin[2\pi\,\phi(t)]+k_{\mathrm{P}} \sin[2\pi \phi(t)]\,,
\end{equation}
where $k_\mathrm{P}$ is the planet's radial-velocity semi-amplitude. The 2D maps of the transmission spectra in the planet rest frame are shown in the appendix (see Fig.~\ref{fig:contour}). 
%
\section{Results}\label{sec:results}
The presence of an extended and possibly escaping atmosphere containing a significant amount of metastable He would appear as absorption features in the transmission spectrum in the planet's rest frame at the position of the stellar helium triplet. However, as Figure~\ref{sp_transmission} shows, this is not the case: we did not detect any significant absorption feature at the position of the He{\sc i} triplet, either considering the single nights or all transits combined together.

We thus estimated the upper limit $c$ (in F$_{\rm in}$/F$_{\rm out}$) of the He{\sc i} absorption in the transmission spectrum of WASP-80b considering the standard deviation of $T_\mathrm{ave}$ in a spectral region around the helium triplet (10829--10836\,\AA). We then translated it into an effective planetary radius $R_\mathrm{eff}$ as
\begin{equation}
\frac{R_\mathrm{eff}}{R_\mathrm{P}}=\sqrt{\frac{\delta+c}{\delta}}\,,
\end{equation}
where $\delta$ is the transit depth (i.e. $(\frac{R_\mathrm{p}}{R_\star})^2$), and $c$ is the upper limit for the detection of a signal given the measured standard deviation of $\sim$0.85$\%$ at the 95\% confidence level. Finally, we estimated an upper limit for the effective radius of $R_\mathrm{eff}~\sim~$1.14~R$_\mathrm{p}$ at the 95\% confidence level. 

We refined this upper limit by performing an injection and retrieval analysis. The injected model consisted in a Gaussian function with a fixed full width at half maximum (FWHM; computed by convolving the instrumental resolution with the planetary tidally locked rotation), a fixed center (the position of the reddest component of the He{\sc i}\,(2$^3$S) triplet and a variable amplitude. We changed the amplitude of the Gaussian function mimicking a planetary signal until the retrieved absorption had a statistical significance of 2$\sigma$ compared to the continuum noise. In this way, we estimated a more accurate upper limit of c$\sim$0.7$\%$ at the 95\% confidence level, which translates into $R_\mathrm{eff}\sim$1.11~R$_\mathrm{p}$ at the 95\% confidence level.
\section{Modelling}\label{sec:model}
\subsection{Stellar spectral energy distribution}\label{sec:sed}
The population of metastable He{\sc i} in the upper planetary atmosphere is affected by the stellar XUV and UV flux, while the stellar nIR emission controls radiation pressure driving the motion of the escaping He atoms \citep[e.g.][]{oklopcic2018,oklopcic2019,lampon2021,khodachenko2021a}. Therefore, to enable modelling the planetary upper atmosphere and thus attempt to reproduce and explain the non-detection of metastable He{\sc i} in WASP-80b, we estimated the stellar emission in the relevant bands as follows.

Except for the He{\sc i} stellar absorption lines, the nIR stellar emission is purely photospheric. To model it, we employed MARCS models \citep{gustafsson2008}, which account for both atomic and molecular opacities. We considered a stellar effective temperature ($T_{\rm eff}$) of 4150$\pm$100\,K and a surface gravity ($\log{g}$) of 4.5 \citep{triaud2013,triaud2015,gaiadr2}. The $\log{g}$ values derived for WASP-80 and listed in the literature are slightly higher than 4.5, namely 4.6--4.7, but $\log{g}$ has a negligible impact on the nIR emission. Instead, $T_{\rm eff}$, which has the largest impact on the nIR flux, is known with a rather large uncertainty of about 100\,K. Also the stellar radius listed in the literature spans between 0.571 and 0.606\,$R_{\odot}$ \citep{triaud2013,triaud2015,bonomo2017,gaiadr2}. For this reason, we estimated the minimum and maximum nIR flux in the region covered by the He{\sc i} lines by combining the minimum and maximum $T_{\rm eff}$ values (i.e. 4050 and 4250\,K) and stellar radii, respectively. In this way, we obtained a minimum value for the nIR continuum flux around 10830\,\AA\ at the distance of the planet \citep[0.0344\,AU;][]{triaud2015} of $\approx$5923\,erg\,cm$^{-2}$\,s$^{-1}$\,\AA$^{-1}$ and a maximum flux of $\approx$7686\,erg\,cm$^{-2}$\,s$^{-1}$\,\AA$^{-1}$. Therefore, as a consequence of the uncertainties on stellar radius and $T_{\rm eff}$, the nIR flux varies by at most a factor of $\approx$1.3.

The XUV emission of WASP-80 has been estimated by \citet{salz2016} and \citet{king2018} on the basis of X-ray observations collected with ROSAT and XMM-Newton. Our starting point for estimating the stellar XUV flux is the result of \citet{king2018}, who derived an XUV flux at the distance of the planet of 8900$\pm$4300\,erg\,cm$^{-2}$\,s$^{-1}$ integrating over the 13.6\,eV and 2.4\,keV range (i.e. 5.2--912\,\AA). The uncertainty on the XUV emission given by \citet{king2018} is dominated mostly by the rather large uncertainty on the stellar distance of 60$\pm$20\,pc, which was based on a photometric parallax. In the meantime, the Gaia satellite provided a significantly more precise distance to the star of 49.73$\pm$0.05\,pc \citep{gaiaedr3_1,gaiaedr3_2}. This enabled us to improve the precision of the measurement of the X-ray flux based on the XMM-Newton observations (L$_{\rm X}$\,=\,4.85$^{+0.12}_{-0.23}\times10^{27}$\,erg\,s$^{-1}$) and thus also the accuracy of the XUV flux, which we estimated employing the scaling relations given by \citet{king2018} and the updated stellar X-ray luminosity. In this way, we obtained an XUV flux at the distance of the planet integrated over the 5.2--912\,\AA\ range of $\approx$6281\,erg\,cm$^{-2}$\,s$^{-1}$, which corresponds to a value of $\approx$7.5\,erg\,cm$^{-2}$\,s$^{-1}$ at the distance of 1\,AU. This value is a factor of about three lower than that obtained using the scaling relation of \citet{sreejith2020} and the measured \logR\ value of about $-$4.04 (see below), which is within the uncertainties.

\begin{figure}[ht!]
\begin{center}
\includegraphics[width=\hsize]{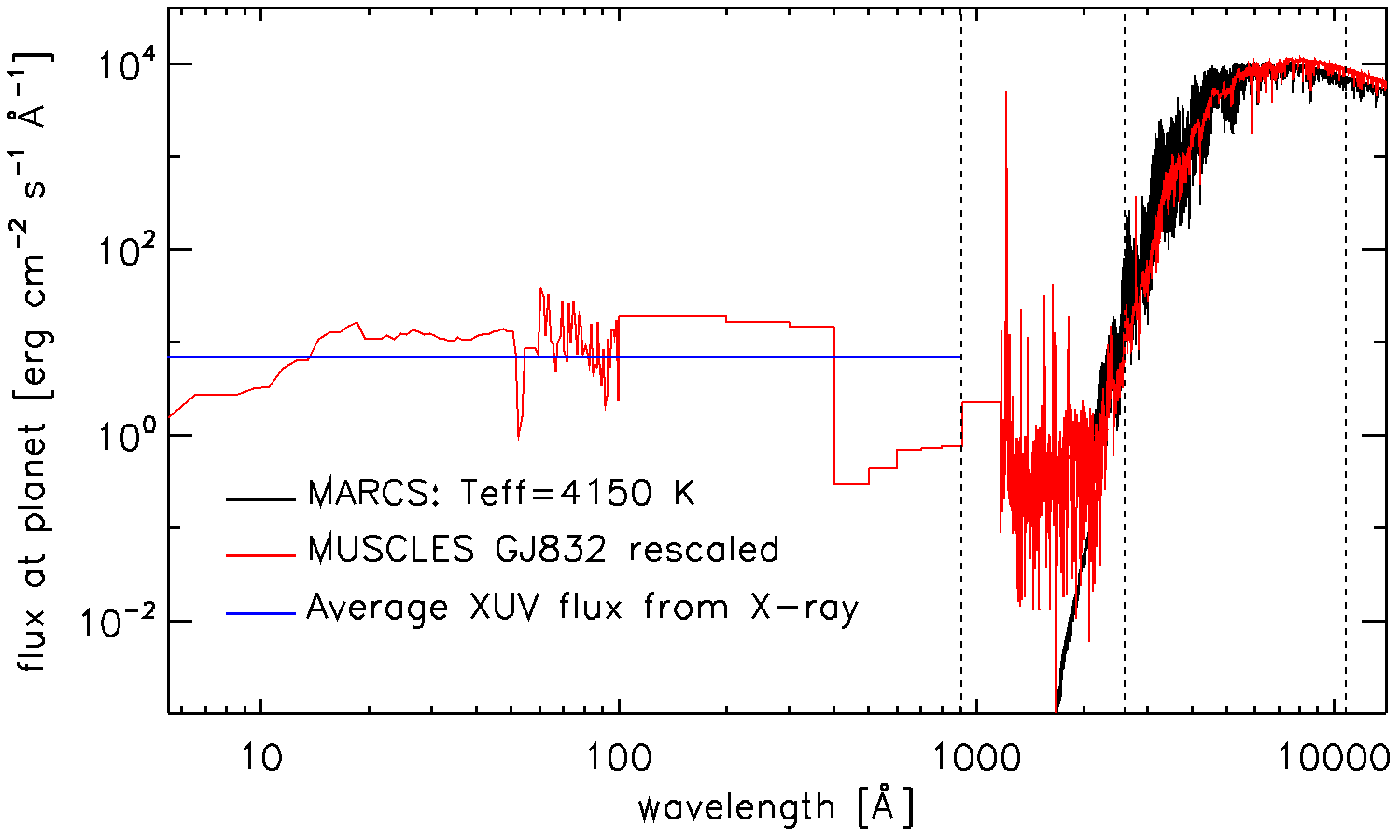}
\caption{Photospheric emission of WASP-80 obtained from MARCS models (black line) and MUSCLES SED of GJ832 (red line) rescaled in such a way to reproduce the XUV emission of WASP-80 derived from the X-ray measurements and the scaling relations of \citet{king2018}. The MARCS model has been convolved to a resolution comparable to that of the MUSCLES spectrum of GJ832 in the optical and infrared. The blue line shows the average XUV stellar emission at the distance of the planet. For reference, the vertical dashed lines indicate from left to right the position of the hydrogen ionisation threshold ($\approx$912\,\AA), the position of the metastable He{\sc i} ionisation threshold ($\approx$2600\,\AA), and the approximate position of the He{\sc i} features ($\approx$10830\,\AA).}
\label{fig:sed}
\end{center}
\end{figure}

The UV (FUV and NUV) emission of WASP-80, particularly at wavelengths below 2600\,\AA, is not photospheric and thus it cannot be estimated employing the MARCS model. Therefore, to estimate the UV flux at the distance of the planet we looked for a star as similar as possible to WASP-80 in terms of both atmospheric parameters and activity and with an observed UV spectrum. GJ832 is an early M dwarf with an effective temperature of about 3600\,K \citep[e.g.][]{kuznetsov2019} and a measured \logR\ value of about $-$5.1 \citep{jenkins2006,boro2018,hojjatpanah2019,sreejith2020}, thus slightly cooler and less active than WASP-80, but with a measured UV flux and a modelled XUV flux that can be used as anchor \citep{france2016}. We estimated the UV spectral emission of WASP-80 by rescaling the stellar flux of GJ832 until the XUV flux matched that derived for WASP-80. In particular, to match the XUV flux of WASP-80, we had to multiply the XUV flux of GJ832 by a factor of 6.8, which is in line with the fact that WASP-80 is more active than GJ832, as indicated by the respective \logR\ values. The rescaled UV spectrum of GJ832 is shown in Figure~\ref{fig:sed} in comparison to the MARCS model and the average XUV flux.

\begin{figure}[ht!]
\begin{center}
\includegraphics[width=\hsize]{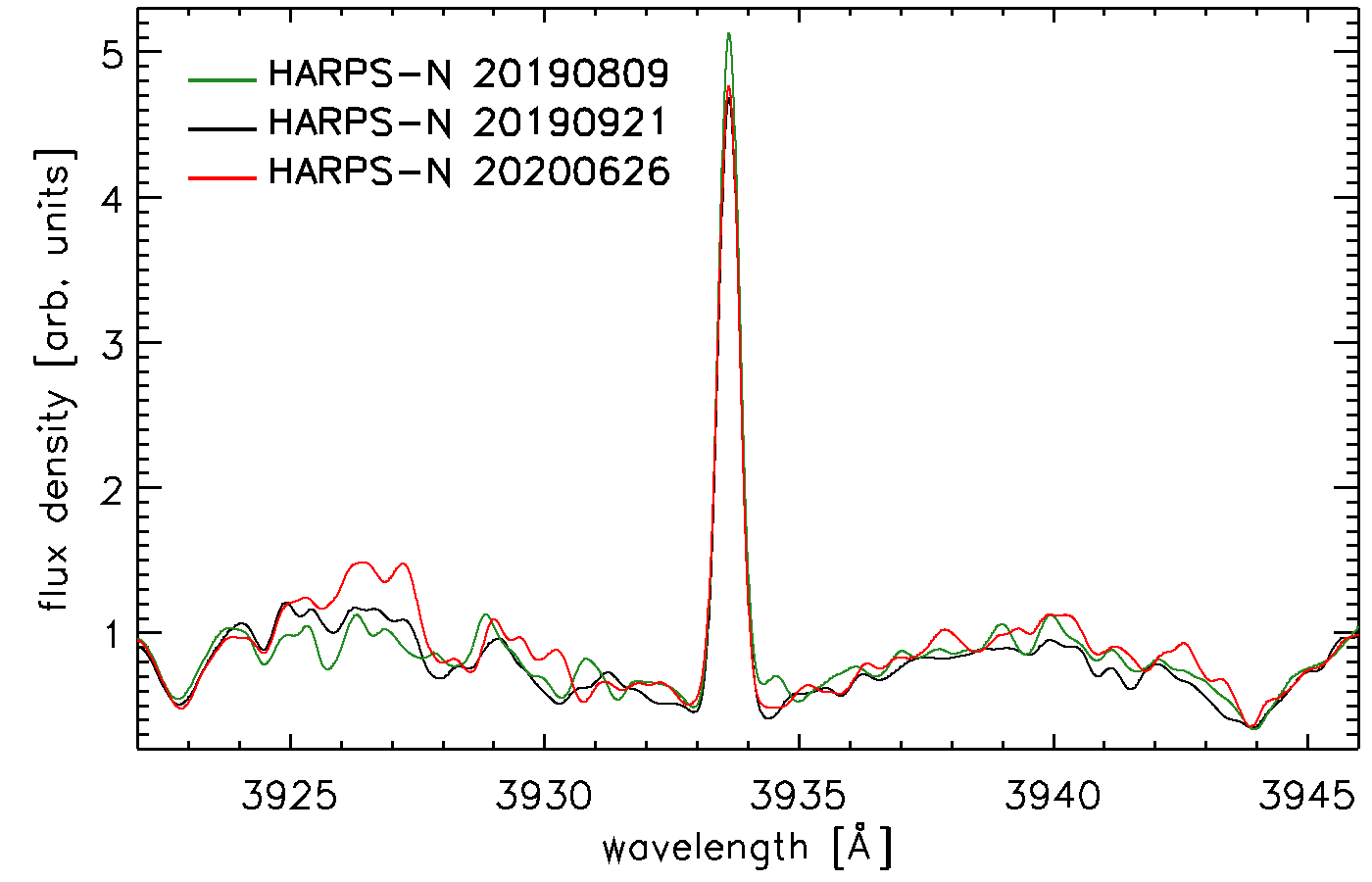}
\caption{Comparison among the master HARPS spectra of the Ca{\sc ii}\,K line obtained combining the data collected within each night and convolved to a spectral resolution of 10,000, for visualisation purposes.}
\label{fig:Ca2comparison}
\end{center}
\end{figure}

Given that WASP-80 is a rather active star, we looked for possible time variations of the stellar high-energy emission by looking at the Ca{\sc ii}\,H\&K line core emission. In particular, we summed up the HARPS-N spectra, obtained simultaneously to the GIANO data, within each night to obtain master spectra to increase the S/N. Figure~\ref{fig:Ca2comparison} shows that the high-energy emission measured from the HARPS-N spectra is comparable among the three nights of our observations. We employed the master spectra to measure \logR\ \citep{noyes1984,rutten1984,fossati2017a} obtaining values ranging between $-$4.02 and $-$4.06. We also looked for variations of the \logR\ value within each night obtaining that the variability lies within the uncertainties. 
%
\subsection{3D hydrodynamic modelling}
The 3D modelling results presented here are based on a multi-fluid self-consistent aeronomy model of the planetary wind and of its interaction with the stellar wind. The code has already been employed to model the upper atmosphere and reproduce the observations of the hot Jupiter HD209458b \citep{ildar2020a}, the warm Neptunes GJ436b \citep{khodachenko2019} and GJ3470b \citep{ildar2021}, and the super-Earth $\pi$\,Men\,c \citep{ildar2020b}. In particular, the works of \citet{ildar2021} and \citet{khodachenko2021a} focus on the modelling of metastable He{\sc i} in the planetary atmospheres and on fitting the relative observations.
\begin{figure*}[ht!]
\begin{center}
\includegraphics[width=8.7cm]{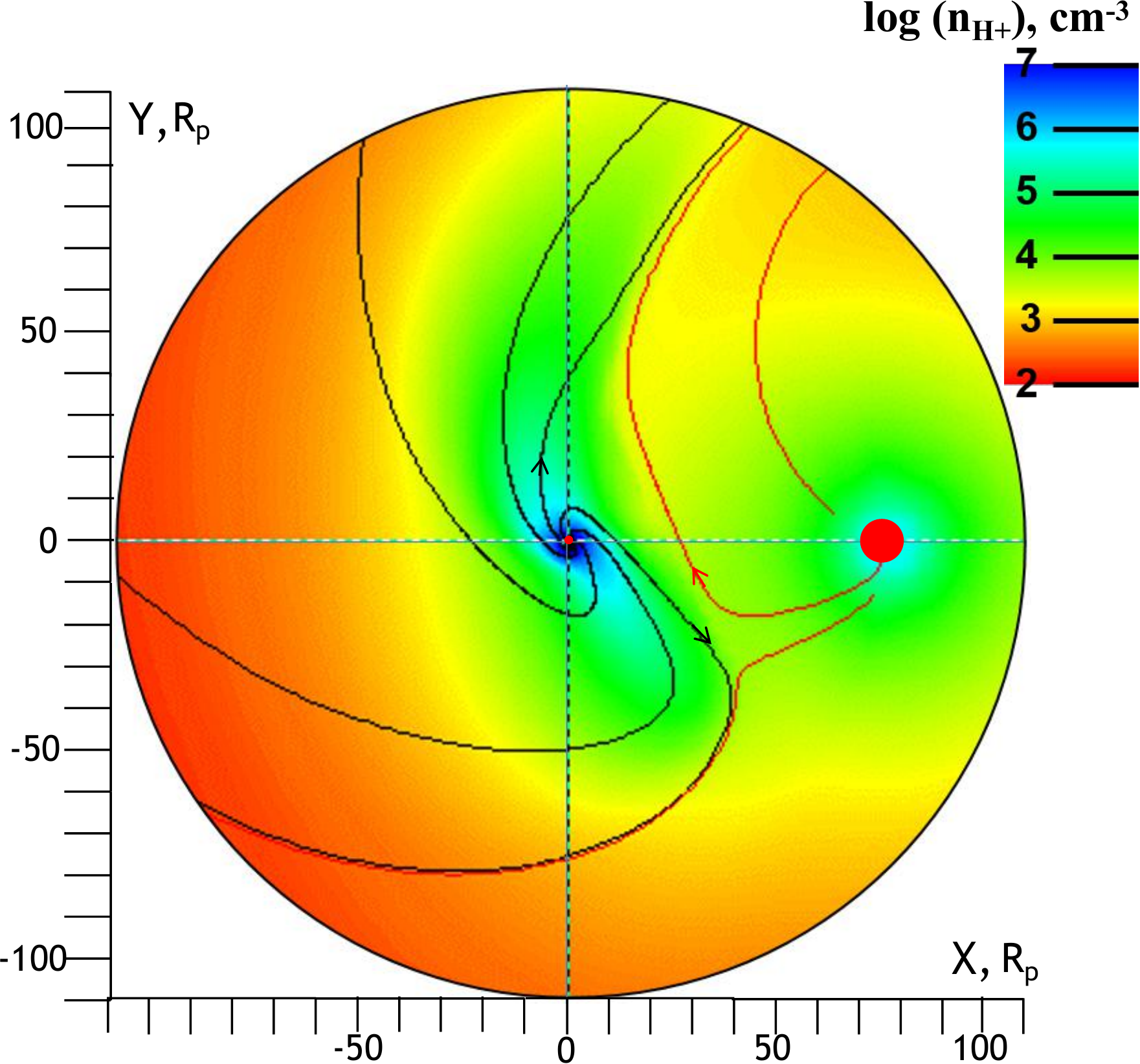}
\includegraphics[width=9.5cm]{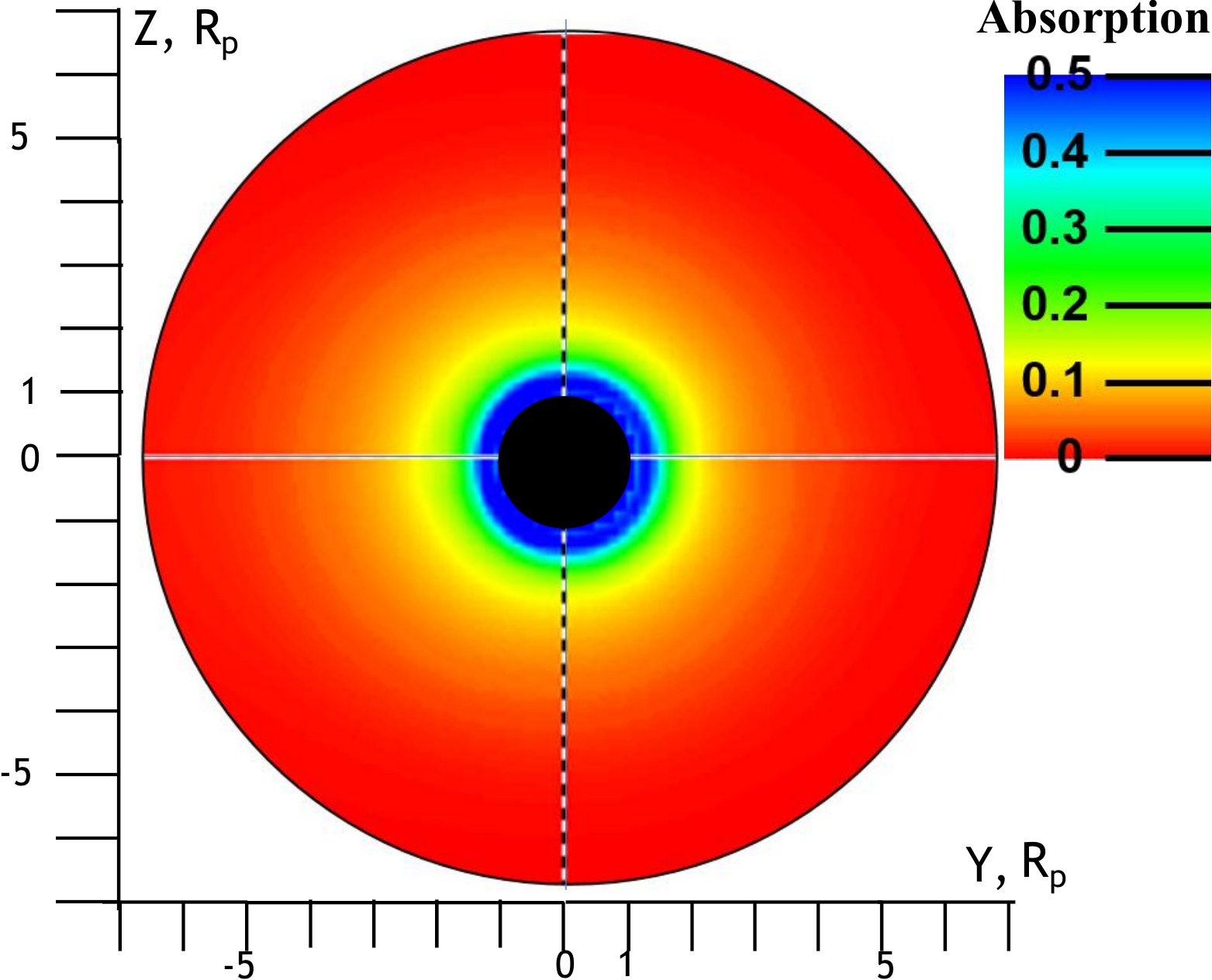}
\caption{Left: proton density distribution in the orbital plane of the whole simulated space comprising the WASP-80 system. The planet is at the center of the coordinate system (0,0) and moves anti-clockwise relative to the star that is located at (76,0). The red dots indicate the position and size of the star (right) and planet (center). The black and red lines correspond to the proton fluid streamlines originated from the planet and the star, respectively. The axes are in units of planetary radii. Right: distribution of metastable He{\sc i} local absorption (from 0, that is no absorption, to 1, that is full absorption) along the line of sight at mid-transit integrated over $\pm$10\,km\,s$^{-1}$. The whole plotted circle corresponds to the stellar disk, while the lower boundary of the simulated planetary atmosphere is shown by the black circle.}
\label{fig:3Dsim}
\end{center}
\end{figure*}
\begin{figure*}[h]
\begin{center}
\includegraphics[width=9.7cm]{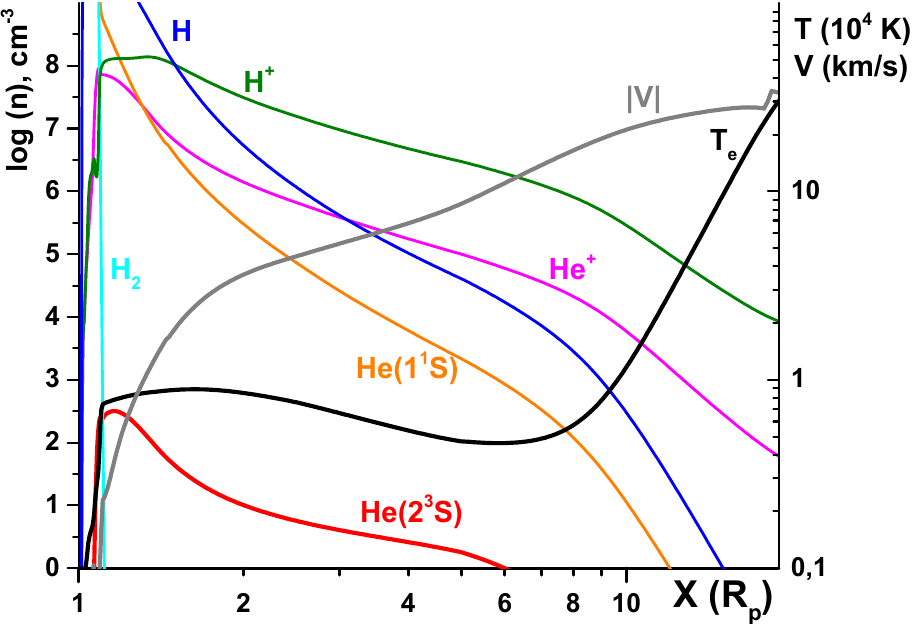}
\includegraphics[width=8.5cm]{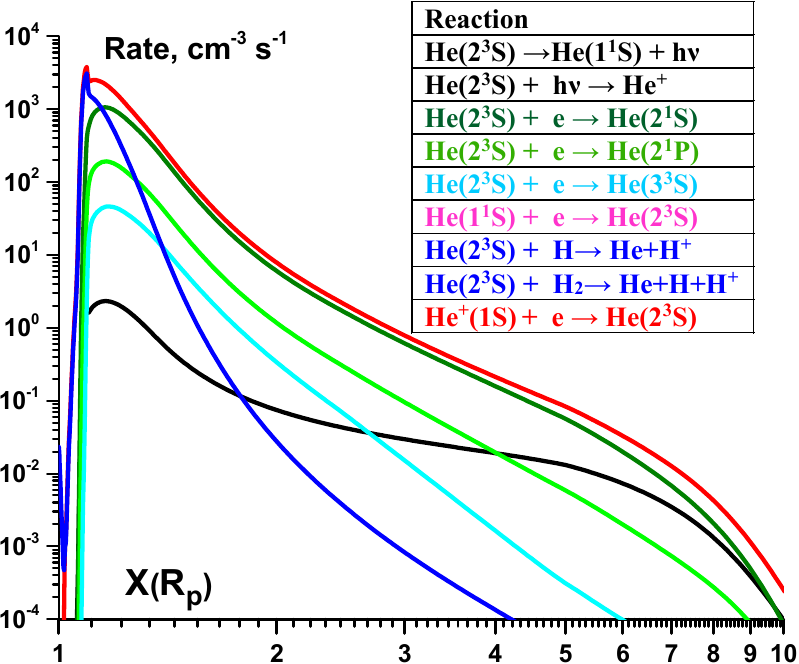}
\caption{Profiles along the planet-to-star direction. Left: temperature (black line) and velocity (gray line) of protons (in 10$^4$\,K and km\,s$^{-1}$, respectively; the scale for both is on the right axis), and density of major species labelled in the plot (left axis). Right: rates of the reactions responsible for the processes of population and depopulation of metastable He{\sc i} listed in the legend. The black line shows the sum of reactions 1 and 2, while the blue line is the sum of reactions 7 and 8 (see legend). The x-axes are in units of planetary radii.}
\label{fig:3Dprofiles}
\end{center}
\end{figure*}

The code solves continuity, momentum, and energy equations for all considered species, which are H, H$^+$, H$_2$, H$_2^+$, H$_3^+$, He, He$^+$, and He$_2^+$. The metastable He{\sc i}\,(2$^3$S) atoms are treated as a separate fluid with its own velocity and temperature, which are determined by those of the species from which they originate, namely He$^+$ or He{\sc i}\,(1$^3$S), depending on whether recombination or excitation from the ground state, respectively, generate the He{\sc i} in the metastable state. Elastic collisions with other species also affect the macroscopic physical parameters of the He{\sc i}\,(2$^3$S) fluid. All reactions, which populate and depopulate the He{\sc i}\,(2$^3$S) component, are those listed by \citet{oklopcic2018} and \citet[][Table 1]{ildar2021}, while the details of the absorption calculations are described in \citet{ildar2021}.
\begin{figure*}[ht!]
\begin{center}
\includegraphics[width=9.1cm]{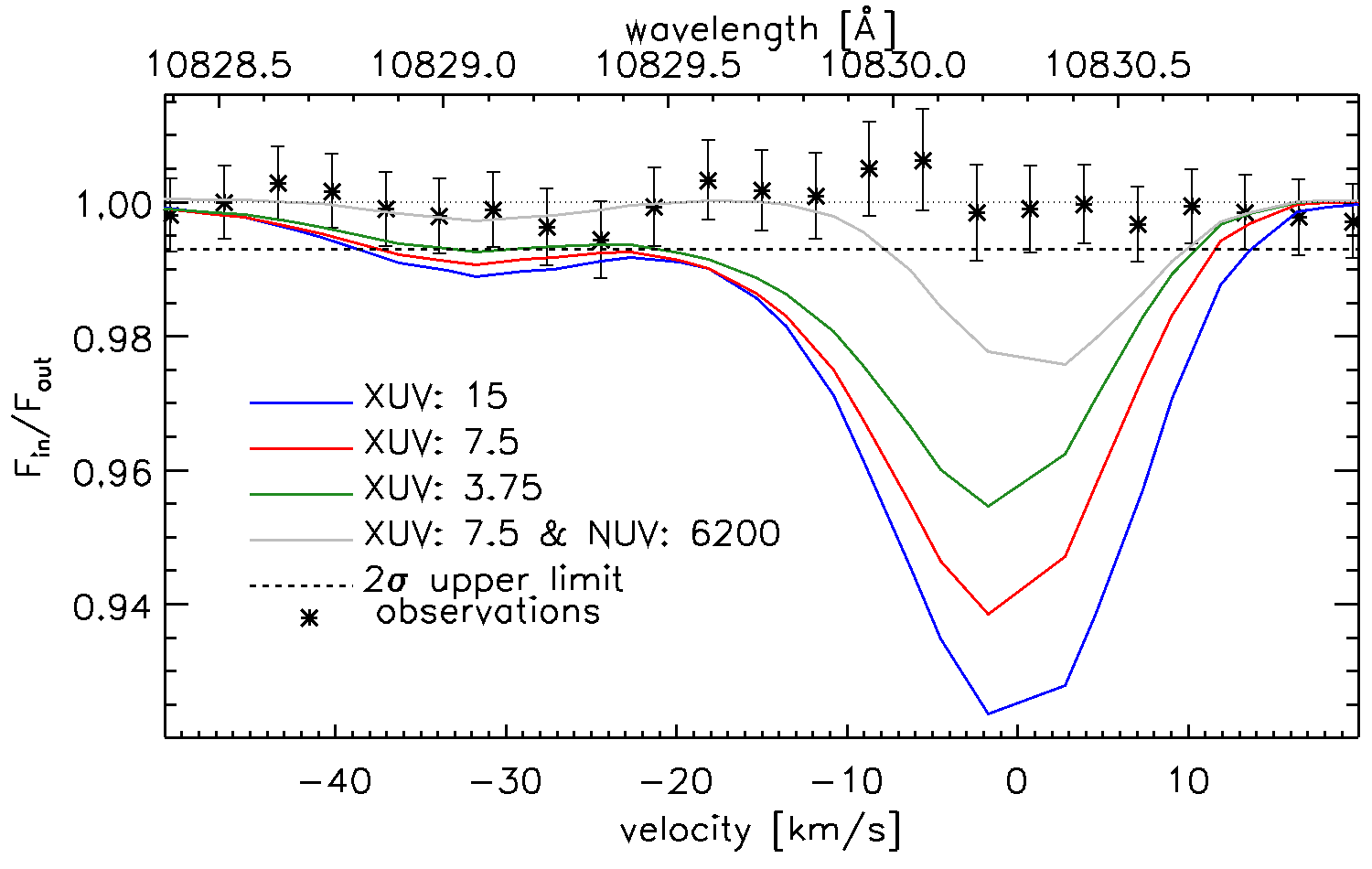}
\includegraphics[width=9.1cm]{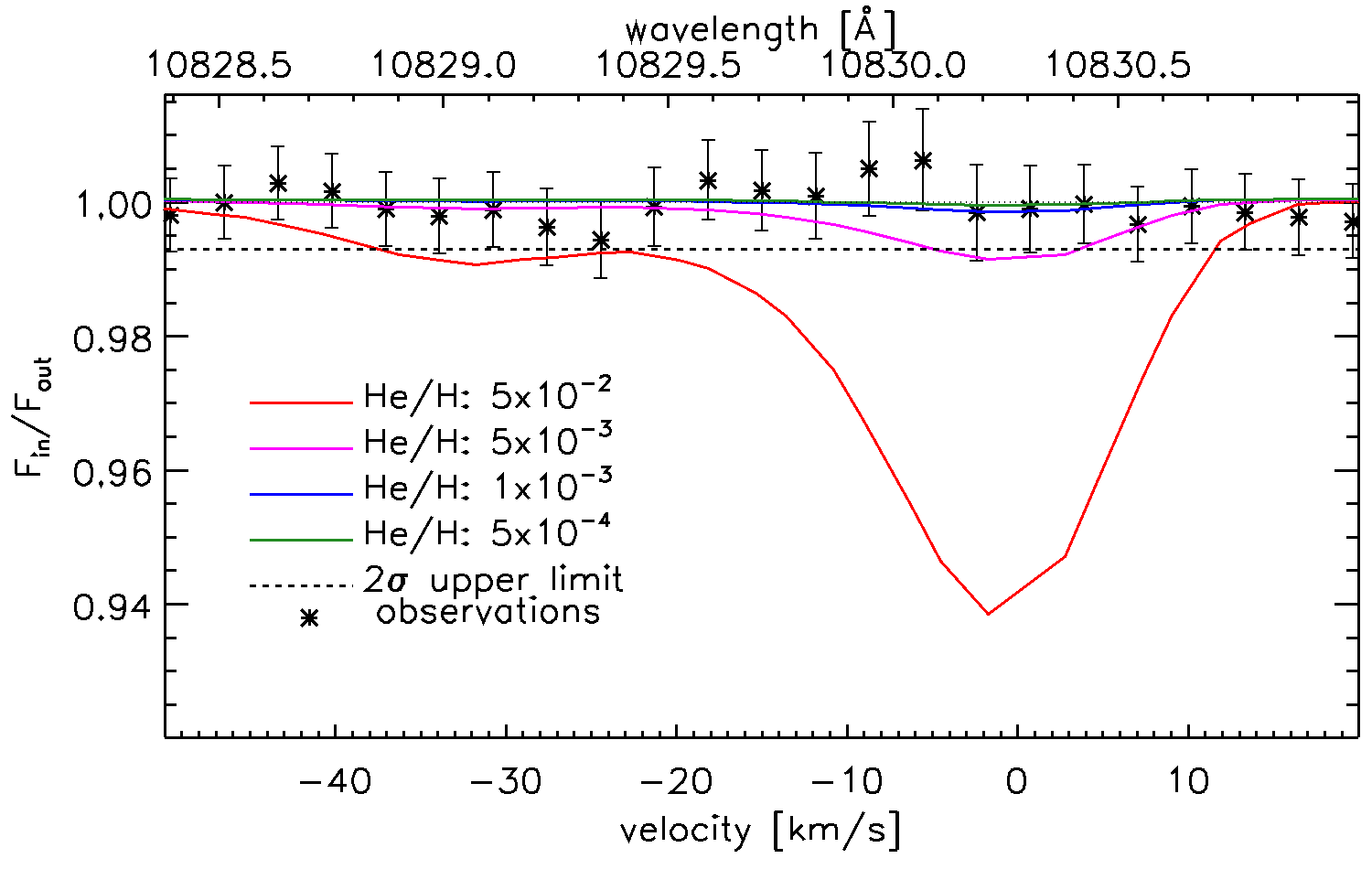}
\caption{Left: He{\sc i}\,(2$^3$S) triplet absorption profiles obtained considering three different values of the stellar XUV flux (blue, red, and green lines) at a fixed He abundance of He/H\,=\,0.05 (the fluxes in the legend are at 1\,AU and in erg\,cm$^{-2}$\,s$^{-1}$) and a 100 times higher value of the stellar NUV flux (gray line). Right: He{\sc i}\,(2$^3$S) triplet absorption profiles obtained considering different values of the He abundance. In both panels, the zero Doppler-shifted velocity on the x-axis corresponds to a wavelength of 10830.25\,\AA, the observed transmission spectrum is shown by black asterisks, and the horizontal dashed line marks the 2$\sigma$ upper limit derived from the observations. The horizontal dotted line at one is for reference.}
\label{fig:3Dabsorption}
\end{center}
\end{figure*}

We employed the UV flux derived as described in Section~\ref{sec:sed} to compute the photoionisation time of He{\sc i}\,(2$^3$S) atoms, obtaining about 2.3 minutes at the planetary orbit. The XUV stellar flux ionises and heats upper atmospheres through the production of photoelectrons, finally leading to hydrodynamic outflow. Different species interact via elastic, charge-exchange, and Coulomb collisions, which efficiently couple velocities and temperatures of atoms and ions in the region dominated by the planetary material \citep[e.g.][]{debrecht2020}. The 3D model also calculates self-consistently the stellar wind plasma over the whole star-planet system. Unless otherwise stated, for the simulations we considered a stellar wind velocity of 200\,km\,s$^{-1}$, a stellar wind temperature of 0.7\,MK, and a stellar wind density of 10$^3$\,cm$^{-3}$ at the position of the planet. These values lead to a stellar mass-loss rate of 10$^{11}$\,g\,s$^{-1}$, namely a sub-solar wind strength, but we test the impact of the stellar wind and present the results later in this section.

The model equations are solved on a spherical grid in the planet-centered reference frame with the polar-axis Z perpendicular to the orbital plane \citep[see][]{ildar2020a}. For all model runs, we set a temperature of 1000\,K and a pressure of 0.05\,bar at the base of the planetary atmosphere. The chosen lower boundary temperature is close to the planetary equilibrium temperature. The lower boundary pressure was chosen such that all XUV photons are absorbed within the simulation domain, that is above the lower boundary \citep[see][for more details]{ildar2014}. Each simulation has been run continuously for 600 dimensionless times\footnote{This is in units of R$_{\rm p}$/V$_0$, where V$_0$ is the proton velocity at a temperature of 10$^4$\,K, that is 9.07\,km\,s$^{-1}$.}, corresponding to about 18 planetary orbits, and convergence of the solution has been ensured by checking the stability of the integral mass loss, which reaches a quasi-stationary level after about 200 dimensionless times (about 6 planetary orbits).

Figure~\ref{fig:3Dsim} presents the structure of the expanding upper planetary atmosphere on the scale of the whole simulation domain obtained assuming a nearly solar He abundance (He/H\,=\,0.05). It shows that the planetary atmosphere extends far from the planet and stretches both ahead and behind the planet along the orbit. From this model, we obtained a total mass-loss rate of 2.6$\times$10$^{10}$\,g\,s$^{-1}$. The simulation shown in Figure~\ref{fig:3Dsim} led to significant He{\sc i}\,(2$^3$S) absorption of the order of 10\%, in strong contrast with the observations. Figure~\ref{fig:3Dsim} shows that most of the absorption takes place relatively close to the planet, within a spherical shell with a radius of about 3\,$R_{\rm p}$.

The left panel of Figure~\ref{fig:3Dprofiles} shows the density, velocity, and temperature profile of the planetary gas gathered from the simulation. As obtained by \citet{salz2016}, molecular hydrogen is dissociated very rapidly and the temperature rises steeply following H$_2$ dissociation reaching temperatures of about 10$^4$\,K, a velocity of about 10\,km\,s$^{-1}$, and a mass-loss rate of 2.6$\times$10$^{10}$\,g\,s$^{-1}$. This is similar to what is found from simulations of classical hot Jupiters \citep[e.g.][]{salz2016,kubyshkina2018,ildar2020a}. The right panel of Figure~\ref{fig:3Dprofiles} shows the reaction rates of kinetic processes populating and depopulating metastable He{\sc i}. The recombination of He$^+$ into He{\sc i}\,(2$^3$S) is balanced by auto-ionisation collisions with H{\sc i} at low altitudes ($<$1.2\,$R_{\rm p}$) and by electron collisional depopulation at higher altitudes. Photoionisation of metastable He{\sc i}, instead, becomes relevant relatively far from the planet, at a distance at which the He{\sc i}\,(2$^3$S) density is too low to affect the absorption during transit. Therefore, the stellar UV flux has a relatively little impact on reproducing the observed non-detection. Similarly, we obtain that radiation pressure accelerating the metastable He{\sc i} atoms has a small impact on the absorption features.

Given the low photoionisation rate of the metastable He{\sc i}, the only parameters affecting the absorption observed during transit are the stellar XUV flux, the He abundance, and possibly the stellar wind strength. Figure~\ref{fig:3Dabsorption} shows the absorption profiles obtained varying the stellar XUV flux and the He abundance in the planetary upper atmosphere. As expected, the absorption depth varies significantly with varying both XUV flux and atmospheric He abundance. In particular, we varied the XUV flux by a factor of two, but this did not enable us to reproduce the observed non-detection. We remark that, as a result of the geometry of the model, the simulations take into account planetary rotation that is assumed to be tidally locked with the planetary orbit around the host star. Therefore, we lowered the atmospheric He abundance until matching the observations obtaining that, for the estimated stellar XUV flux, the absorption depth falls below 0.7\% for He/H values smaller than 5$\times$10$^{-3}$, that is about 16 times smaller than solar, thus putting a strong constraint on the He abundance in the planetary upper atmosphere. To strengthen this result, we computed a further model increasing the stellar UV flux, thus the metastable He{\sc i} photoionisation rate, by 100 times, but also in this case, we obtained a significant He{\sc i} absorption in contrast to the observations.

\begin{figure}[h]
\begin{center}
\includegraphics[width=\columnwidth]{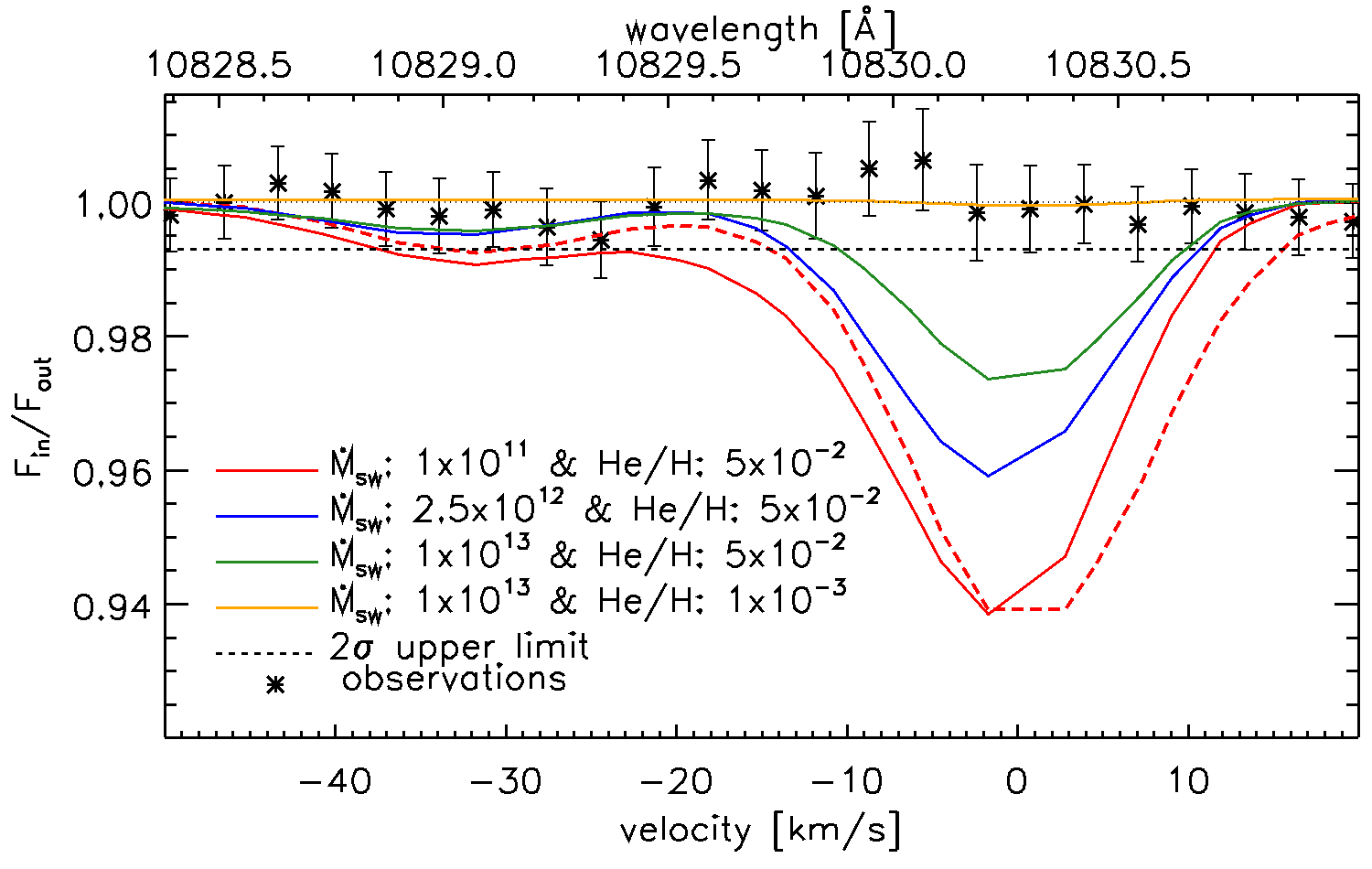}
\caption{He{\sc i} triplet absorption profiles obtained considering a fixed stellar XUV flux at the distance of 1\,AU of 7.5\,erg\,cm$^{-2}$\,s$^{-1}$, a solar He abundance, and different stellar wind strengths expressed in terms of mass-loss rate in g\,s$^{-1}$. Considering the strongest stellar wind taken into account, the observed non-detection is reached for He/H values smaller than $\approx$0.01. The red long dashed line shows the absorption obtained considering the same conditions taken for the solid red line except for a ten times weaker radiation pressure on the He{\sc i} metastable atoms. The zero Doppler-shifted velocity on the x-axis corresponds to a wavelength of 10830.25\,\AA, the observed transmission spectrum is shown by black asterisks, and the horizontal dashed line marks the 2$\sigma$ upper limit. The horizontal dotted line at one that we placed for reference is largely masked by the yellow line.}
\label{fig:SW}
\end{center}
\end{figure}

A further parameter possibly affecting the He{\sc i} absorption signal is the stellar wind, which can cause the planetary atmosphere to compress \citep[e.g.][]{vidotto2020}. Figure~\ref{fig:SW} shows how the stellar wind affects the He{\sc i} absorption. Indeed, a stronger stellar wind compared to what was used for the simulations shown above reduces the He{\sc i} absorption by about 1.5 times. This is the result of the stellar wind compressing the planetary atmospheric outflow with the bow-shock moving as close as 3.5 planetary radii. As a result, the atmospheric gas responsible for the absorption remains close to the planet, reducing the absorption signal. Figure~\ref{fig:SW} shows that an even stronger stellar wind with a mass-loss rate of 10$^{13}$\,g\,s$^{-1}$ leads to the bow-shock moving at about 2.2 planetary radii, further reducing the He{\sc i} absorption signal. Therefore, we obtain that with the strongest stellar wind we consider, which is about four times stronger than solar and 16 times stronger than that derived for the K-type star HD219134 \citep{vidotto2018}, the He abundance leading to fit the non-detection is He/H\,$\lesssim$\,0.01, that is about a factor ten smaller than solar.

Figure~\ref{fig:SW} shows also the effect that radiation pressure plays on the He{\sc i} absorption. Artificially reducing the radiation pressure impinging on the metastable He{\sc i} atoms by ten times shifts the absorption profile by about 2.5\,km\,s$^{-1}$ towards the red.
\section{Discussion and conclusion}\label{sec:dicsconc}
We employed the GIANO-B spectrograph to obtain the transmission spectrum of the hot Jupiter WASP-80b in the wavelength range covered by the triplet of metastable He{\sc i} at 10830\,\AA. In particular, we looked for He{\sc i} absorption indicative of the presence of an extended and possibly escaping atmosphere. As a matter of fact, hydrodynamic modelling and considerations based on the planetary properties suggest that WASP-80b should be a prime target for the detection of metastable neutral He in the planetary upper atmosphere. However, the GIANO-B observations led to a non-detection, with an upper limit of 0.11\,$R_{\rm p}$ (95\% confidence level) on the size of the possible He{\sc i} absorption signal.

We attempted to understand this unexpected result employing a 3D aeronomy model that has proven to be working well in reproducing the He{\sc i} transmission spectroscopy observations of GJ3470b and WASP-107b \citep{ildar2021,khodachenko2021a}, as well as of HD209458b and HD189733b (in prep.). Even the presence of high-altitude clouds would not aid explaining the non-detection. Indeed, considering a solar He abundance, the peak of the density of metastable He lies close to the layers absorbing most of the stellar XUV photons (Figure~\ref{fig:3Dprofiles}), where no clouds can form because of photodissociation. Given the system properties, the most plausible solution suggested by the model for reproducing the non-detection is a low He abundance. In particular, we find He/H\,$<$\,5$\times$10$^{-3}$ (about 16 times smaller than solar) for a stellar wind 25 times weaker than solar or He/H\,$<$\,10$^{-2}$ (about ten times smaller than solar) for a stellar wind four times stronger than solar. This is a remarkable result considering that for WASP-107b, which has physical properties similar to those of WASP-80b, the same model returns a solar He abundance. The major difference between the two planets is a factor five higher mass of WASP-80b compared to WASP-107b, which leads to a more compact atmosphere for WASP-80b.

We attempt to understand the non-detection of He{\sc i} in the upper atmosphere of WASP-80b by putting it in the wider context of published detections and non-detections. To this end, we collected the physical properties of the systems for which either measurements or non-detections of the He{\sc i} metastable triplet have been reported (Table~\ref{tab:systems}). We collected the system parameters from the literature giving priority to more recent and/or homogeneous sources. The stellar XUV fluxes listed in Table~\ref{tab:systems} have been either taken from the literature or, when unavailable, extracted employing the scaling relations of \citet{sreejith2020}. The XUV fluxes listed in Table~\ref{tab:systems} correspond to the amount of high-energy stellar radiation at wavelengths shorter than 912\,\AA\ irradiating the planet. This is not the most relevant wavelength range\footnote{The most relevant wavelength range would be that at wavelengths shorter than 500\,\AA.} in relation to the production of metastable He{\sc i}, but it is still representative of the high-energy stellar emission, it is the one typically reported in the literature, and it can be readily estimated on the basis of a variety of measurements obtained at X-ray, ultraviolet, and optical wavelengths \citep[e.g.][]{Sanz-Forcada2011,Chadney2015,king2018,linsky2013,linsky2014,france2018,sreejith2020}.
\begin{sidewaystable*}
\caption{Properties of the systems for which either measurements or non-detections of the He{\sc i} metastable triplet have been published.}
\label{tab:systems}
\begin{center}
\begin{tabular}{l|ccc|cccc|cc|c}
\hline
\hline
Planet & $T_{\rm eff}$ & $R_{\rm s}$   & XUV flux                   & $M_{\rm p}$   & $R_{\rm p}$   & a    & $T_{\rm eq}$ & $\left(\frac{R_{\rm p}}{R_{\rm s}}\right)^2$ & $\delta_{\rm HeI}$ & Upper \\
       & [K]           & [$R_{\odot}$] & [erg\,cm$^{-2}$\,s$^{-1}$] & [$M_{\rm J}$] & [$R_{\rm J}$] & [AU] & [K]          &                               & [$R_{\rm p}$]      & limit \\
\hline
WASP-80b    & 4150$\pm$100$^1$    & 0.586$\pm$0.018$^2$    &  6281$^3$          & 0.538$\pm$0.036$^2$      & 0.999$\pm$0.031$^2$      & 0.0344$^2$     &  816$\pm$20  & 0.03127 & 0.11$^3$       & Y \\
HD209458b   & 6065$\pm$50$^1$     & 1.178$\pm$0.009$^4$    &  2407$^5$          & 0.682$\pm$0.015$^1$      & 1.359$\pm$0.019$^1$      & 0.04707$^1$    & 1463$\pm$12  & 0.01345 & 0.29$^5$       & N \\
HD189733b   & 5040$\pm$50$^1$     &  0.78$\pm$0.02$^4$     & 28095$^6$          & 1.123$\pm$0.045$^1$      & 1.138$\pm$0.027$^1$      & 0.031$^1$      & 1219$\pm$13  & 0.02152 & 0.16$^{7,8}$   & N \\
WASP-107b   & 4425$\pm$70$^9$     &  0.67$\pm$0.02$^9$     &  6323$^{10}$       & 0.096$\pm$0.005$^9$      &  0.94$\pm$0.02$^{11}$    & 0.055$^{11}$   &  757$\pm$12  & 0.01990 & 0.99$^{12,13}$ & N \\
WASP-69b    & 4700$\pm$50$^1$     & 0.818$\pm$0.025$^4$    &  6192$^{14,15,16}$ & 0.250$\pm$0.023$^1$      & 1.057$\pm$0.047$^1$      & 0.04527$^1$    &  963$\pm$11  & 0.01688 & 0.77$^{16}$    & N \\
GJ436b      & 3479$\pm$60$^{17}$  & 0.449$\pm$0.019$^{17}$ &  2428$^{18}$       & 0.080$\pm$0.007$^{19}$   & 0.366$\pm$0.014$^{19}$   & 0.0308$^{19}$  &  641$\pm$11  & 0.00672 & 0.27$^{16}$    & Y \\
KELT9b      & 9600$\pm$400$^{20}$ & 2.418$\pm$0.058$^{20}$ &  3174$^{21}$       & 2.88$\pm$0.35$^{20}$     & 1.936$\pm$0.047$^{20}$   & 0.03368$^{20}$ & 3922$\pm$165 & 0.00648 & 0.23$^{16}$    & Y \\
GJ3470b     & 3652$\pm$50$^{22}$  & 0.48$\pm$0.04$^{22}$   &  5621$^{23}$       & 0.040$\pm$0.004$^{22}$   & 0.346$\pm$0.029$^{22}$   & 0.0348$^{24}$  &  654$\pm$10  & 0.00525 & 0.96$^{25}$    & N \\
GJ9827b     & 4340$\pm$50$^{26}$  & 0.647$\pm$0.08$^4$     & 36810$^{27}$       & 0.0154$\pm$0.0015$^{26}$ & 0.1407$\pm$0.0028$^{26}$ & 0.0188$^{26}$  & 1228$\pm$26  & 0.00048 & 0.76$^{27}$    & Y \\
GJ9827d     & 4340$\pm$50$^{26}$  & 0.647$\pm$0.08$^4$     &  4162$^{27}$       & 0.0127$\pm$0.0026$^{26}$ & 0.1804$\pm$0.0041$^{26}$ & 0.0559$^{26}$  &  712$\pm$11  & 0.00079 & 1.94$^{27}$    & Y \\
HAT-P-11b   & 4780$\pm$50$^{28}$  & 0.769$\pm$0.048$^4$    &  3236$^{29}$       & 0.0736$\pm$0.0047$^{30}$ & 0.389$\pm$0.005$^{30}$   & 0.05254$^{30}$ &  882$\pm$11  & 0.00259 & 1.27$^{31,32}$ & N \\
HAT-P-18b   & 4803$\pm$80$^1$     & 0.73$\pm$0.04$^4$      &  8000$^{33}$       & 0.200$\pm$0.019$^1$      & 0.995$\pm$0.052$^1$      & 0.05596$^1$    &  837$\pm$14  & 0.01878 & 0.12$^{33}$    & N \\
55 Cnc e    & 5196$\pm$24$^{34}$  & 0.95$\pm$0.08$^4$      &  7413$^{29}$       & 0.0251$\pm$0.001$^{35}$  & 0.1673$\pm$0.0026$^{35}$ & 0.01544$^{35}$ & 1965$\pm$33  & 0.00031 & 0.34$^{36}$    & Y \\
GJ1214b     & 3250$\pm$100$^{37}$ & 0.221$\pm$0.004$^{37}$ &   851$^{29}$       & 0.0197$\pm$0.0027$^{38}$ & 0.254$\pm$0.018$^{38}$   & 0.01411$^{38}$ &  621$\pm$19  & 0.01332 & 1.10$^{39}$    & Y \\
HD63433b    & 5640$\pm$74$^{40}$  & 0.912$\pm$0.034$^{40}$ & 22826$^{41}$       & $-$                      & 0.192$\pm$0.009$^{40}$   & 0.0719$^{40}$  &  969$\pm$13  & 0.00045 & 2.49$^{41}$    & Y \\
HD63433c    & 5640$\pm$74$^{40}$  & 0.912$\pm$0.034$^{40}$ &  5551$^{41}$       & $-$                      & 0.2418$\pm$0.0125$^{40}$ & 0.1458$^{40}$  &  680$\pm$9   & 0.00071 & 1.83$^{41}$    & Y \\
WASP-12b    & 6250$\pm$100$^1$    & 1.57$\pm$0.25$^4$      & 13994$^{15,29}$    & 1.39$\pm$0.12$^1$        & 1.825$\pm$0.091$^1$      & 0.02312$^1$    & 2484$\pm$88  & 0.01366 & 0.01$^{42}$    & Y \\
Trappist-1b & 2559$\pm$50$^{43}$  & 0.117$\pm$0.0036$^{43}$&  1674$^{44}$       & 0.0032$\pm$0.0005$^{45}$ & 0.100$\pm$0.003$^{45}$   & 0.01155$^{45}$ &  393$\pm$8   & 0.00739 & 0.04$^{46}$    & Y \\
Trappist-1e & 2559$\pm$50$^{43}$  & 0.117$\pm$0.0036$^{43}$&   260$^{44}$       & 0.0024$\pm$0.0003$^{45}$ & 0.0812$\pm$0.0025$^{45}$ & 0.02928$^{45}$ &  247$\pm$5   & 0.00487 & 0.07$^{46}$    & Y \\
Trappist-1f & 2559$\pm$50$^{43}$  & 0.117$\pm$0.0036$^{43}$&   150$^{44}$       & 0.0029$\pm$0.0002$^{45}$ & 0.0933$\pm$0.0027$^{45}$ & 0.03853$^{45}$ &  215$\pm$4   & 0.00643 & 0.02$^{46}$    & Y \\
WASP-76b    & 6316$\pm$64$^{47}$  & 1.77$\pm$0.07$^{47}$   &  9105$^{15,48}$    & 0.92$\pm$0.03$^{49}$     & 1.83$\pm$0.05$^{49}$     & 0.033$^{49}$   & 2231$\pm$27  & 0.01081 & 0.35$^{48}$    & Y \\
HAT-P-32b   & 6269$\pm$64$^{50}$  & 1.219$\pm$0.016$^{51}$ & 420000$^{52}$      & 0.585$\pm$0.031$^{51}$   & 1.789$\pm$0.025$^{51}$   & 0.0343$^{51}$  & 1816$\pm$19  & 0.02177 & 1.02$^{52}$    & N \\
\hline
\end{tabular}
\tablefoot{The fourth column lists the XUV stellar flux at wavelengths shorter than 912\,\AA\ impinging on the planet. Column ten gives the effective absorption of the He{\sc i} triplet. References: 1. \citet{bonomo2017}; 2. \citet{triaud2015}; 3. This work; 4.\citet{gaiadr2}; 5. \citet{alonso2019}; 6. \citet{bourrier2020}; 7. \citet{salz2018}; 8. \citet{guilluy2020}; 9. \citet{piaulet2021}; 10. \citet{khodachenko2021a}; 11. \citet{anderson2017}; 12. \citet{allart2019}; 13. \citet{kirk2020}; 14. \citet{figueira2014}; 15. \citet{sreejith2020}; 16. \citet{nortmann2018}; 17. \citet{bourrier2018a}; 18. \citet{bourrier2016}; 19. \citet{lanotte2014}; 20. \citet{borsa2019}; 21. \citet{fossati2018}; 22. \citet{kosiarek2019}; 23. \citet{bourrier2021}; 24. \citet{bonfils2012}; 25. \citet{palle2020}; 26. \citet{rice2019}; 27. \citet{carleo2021}; 28. \citet{bakos2010}; 29. \citet{salz2016}; 30. \citet{yee2018}; 31. \citet{allart2018}; 32. \citet{mansfield2018}; 33. \citet{paragas2021}; 34. \citet{vonbraun2011}; 35. \citet{bourrier2018b}; 36. \citet{zhang2021a}; 37. \citet{gillon2014}; 38. \citet{harpsoe2013}; 39. \citet{crossfield2019}; 40. \citet{mann2020}; 41. \citet{zhang2021b}; 42. \citet{kreidberg2018}; 43. \citet{gillon2017}; 44. \citet{bourrier2017}; 45. \citet{grimm2018}; 46. \citet{krishnamurthy2021}; 47. \citet{tabernero2021}; 48. \citet{casasayas2021}; 49. \citet{ehrenreich2020}; 50. \citet{zhao2014}; 51. \citet{hartman2011}; 52. \citet{czesla2021}} 
\end{center}
\end{sidewaystable*}

\begin{figure*}[h]
\begin{center}
\includegraphics[width=\hsize]{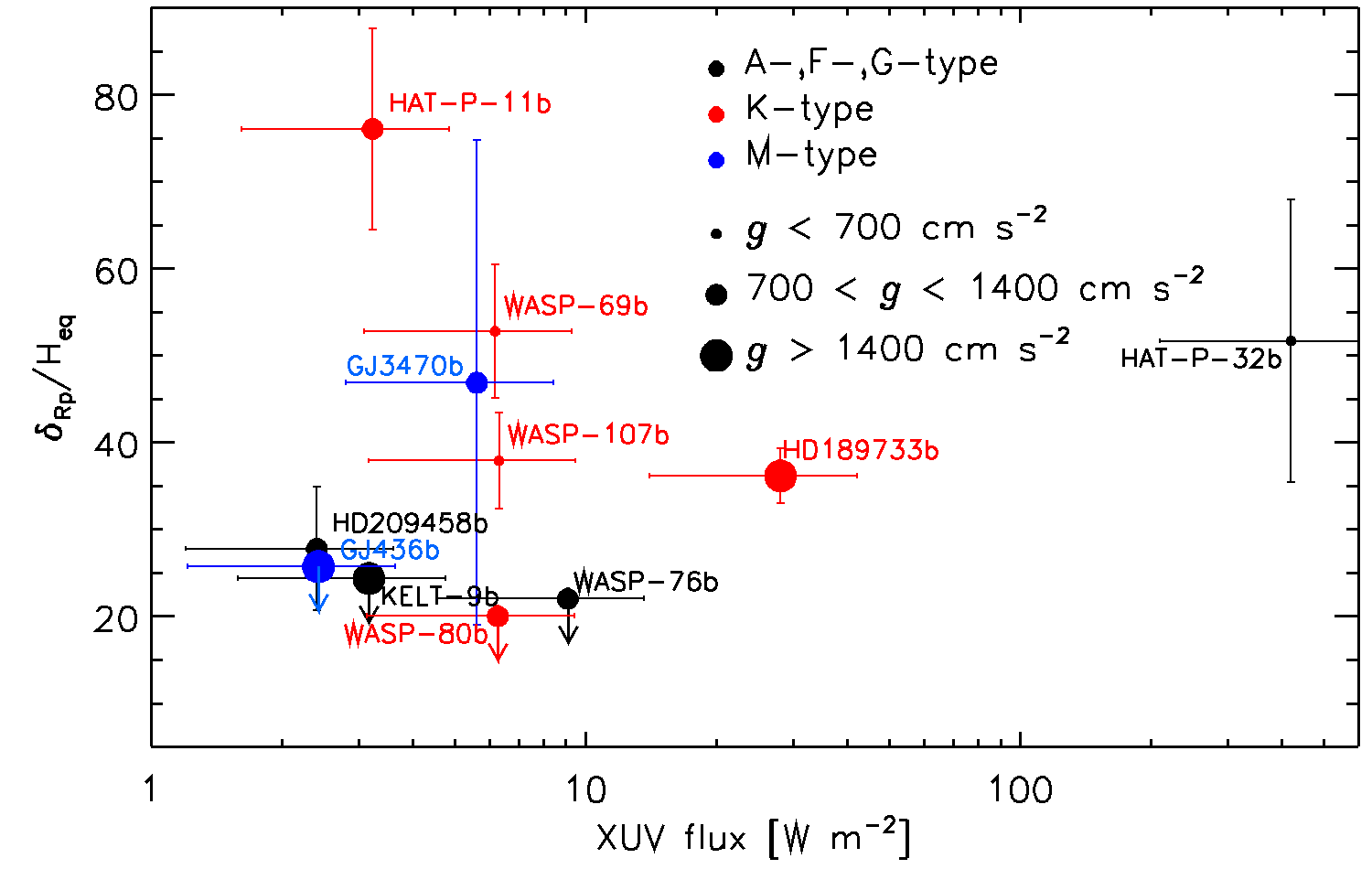}
\caption{Size of the measured He{\sc i} absorption signal, normalised to the atmospheric scale height computed considering the planetary parameters listed in Table~\ref{tab:systems} and a mean molecular weight of a pure hydrogen atmosphere, as a function of the incident stellar XUV flux (in logarithmic scale), with the symbol size indicating the planetary surface gravity. The uncertainties on the stellar XUV flux are set to be equal to a factor of two (see text). Downward arrows indicate upper limits.}
\label{fig:He_alltogether}
\end{center}
\end{figure*}

So far, the search of the He{\sc i} metastable triplet in the upper atmosphere of an exoplanet has been reported for 21 planets ranging from Earth-size rocky planets (e.g. Trappist-1b,e,f) to ultra-hot Jupiters (e.g. KELT-9b), but just for seven of them the observations have led to a positive detection, while in all other cases only an upper limit could be derived. This suggests that the presence of metastable He is not ubiquitous in exoplanetary atmospheres, even when the atmosphere is clearly in a hydrodynamic state (e.g. GJ436b). This is due to the fact that the shape of the stellar spectral energy distribution plays a central role in the production and destruction of metastable He \citep{oklopcic2019}, but other important factors may play a significant role. To attempt uncovering them, we show in Figure~\ref{fig:He_alltogether} the size of the measured He{\sc i} absorption signal or upper limit ($\delta_{\rm Rp}$), normalised to the atmospheric scale height ($H_{\rm eq}$), as a function of incident stellar XUV flux \citep[see also][]{nortmann2018}, with the planetary surface gravity ($g$) indicated by the symbol size (Figure~\ref{fig:He_summary} presents similar plots showing $\delta_{\rm Rp}$/$H_{\rm eq}$ as a function of incident stellar XUV flux and planetary surface gravity, separately). We computed the atmospheric pressure scale height as
\begin{equation}
H_{\rm eq}=\frac{k_{\rm B}\,T_{\rm eq}}{\mu\,g}
\end{equation}
and its uncertainty as
\begin{equation}
\sigma_{H_{\rm eq}}=\sqrt{ \left(\frac{k_{\rm B}}{\mu\,g}\,\sigma_{T_{\rm eq}}\right)^2 + \left(\frac{k_{\rm B}\,T_{\rm eq}}{\mu\,g^2}\,\sigma_{g}\right)^2 }\,,
\end{equation}
where $k_{\rm B}$ is the Boltzmann constant, $T_{\rm eq}$ is the planetary equilibrium temperature listed in Table~\ref{tab:systems}, $g$ is the planetary gravity computed from the planetary mass and radius listed in Table~\ref{tab:systems}, $\mu$ is the mean molecular weight (we assume a hydrogen-dominated atmosphere\footnote{We considered a hydrogen-dominated atmosphere, instead of a hydrogen-and-helium-dominated atmosphere, to reduce the uncertainties given by the unknown He abundance. This choice does not affect the possible detectability of a correlation between $\delta_{\rm Rp}$/$H_{\rm eq}$ and the system parameters.} and hence a value of 1.3 times the mass of a hydrogen atom), and $\sigma_{T_{\rm eq}}$ and $\sigma_{g}$ are the uncertainties on the planetary equilibrium temperature and gravity, respectively. To obtain a more homogeneous sample, we considered only planets with a radius larger than three Earth radii (i.e. we consider only planets that most certainly host a hydrogen-dominated atmosphere). Furthermore, we excluded WASP-12b and HAT-P-18b, whose measurements have been conducted employing low-resolution observations and thus not compatible with the more sensitive high-resolution observations employed for the other reported measurements. Furthermore, following the typical uncertainties of scaling relations employed to convert X-ray, ultraviolet, or optical measurements into XUV flux \citep[e.g.][]{linsky2013,linsky2014,france2018,sreejith2020}, we assigned an uncertainty to the stellar XUV flux of a factor of two. 

Figure~\ref{fig:He_alltogether} does not indicate the presence of any clear correlation with any of the considered parameters (i.e. stellar spectral type, XUV irradiation, and planetary surface gravity), which are those believed to primarily drive the possible detection, or non-detection, of metastable He in planetary upper atmospheres. Although the number of planets for which the detection of the He{\sc i} metastable triplet has been (successfully) attempted is still too small to draw clear conclusions, Figure~\ref{fig:He_alltogether} suggests either the presence of a problem in our understanding of the formation of the He{\sc i} metastable triplet and/or the presence of further parameters playing a significant role in the formation of the triplet. Following the main result obtained from our modelling of WASP-80b and previous results \citep[GJ3470b, HD209458b, HD189733b;][]{ninan2020,palle2020,alonso2019,lampon2020,lampon2021}, this parameter may be the He abundance and/or the stellar wind, though the stellar wind alone does not seem to be capable of reducing enough the planetary absorption signal without a significantly sub-solar He/H abundance ratio, at least in the case of WASP-80b.

There are physical processes that have been identified leading to decrease the He atmospheric abundance with respect to that of hydrogen. A mechanism possibly at work resulting in a reduction of the He abundance in the upper atmospheric layers is phase separation of He and liquid metallic hydrogen for which the former condenses and rains down towards the deeper atmospheric layers \citep[e.g.][]{stevenson1977a,stevenson1977b,guillot2015}. This mechanism should be important for old planets less massive than Jupiter that furthermore remained for a significant amount of time far from the host star before migrating inwards \citep{fortney2004}. Also, magnetic fields may be responsible for the non-detection of He{\sc i} absorption. This is because magnetic fields strongly affect the motion of the atmospheric gas possibly in a way that makes the detection of metastable He{\sc i} impossible with the data at hand \citep[e.g.][]{adams2011,trammel2014,khodachenko2015,khodachenko2021b}.

Unfortunately, it is currently not possible to infer the He abundance of a planetary atmosphere from indicators different than the He{\sc i} triplet. Similarly, it is very difficult to reliably constrain the wind strength of late-type stars in absence of specific indicators such as a Ly$\alpha$ line amenable to reconstruction or the detection of radio emission \citep[e.g.][]{wood2005,fichtinger2017,vidotto2018,folsom2018,folsom2020}. Therefore, it will be important to continue investigating this feature both theoretically and observationally. In particular, it is key that more in-depth modelling is carried out to identify the possible presence of additional physical factors controlling the formation and destruction of metastable He. Also, it is similarly important to attempt the observation of the triplet in other planets, though the focus should be on close-in giant planets, hosting a hydrogen-dominated envelope, to enable comparing results within a larger and yet homogeneous sample of planets. Along the same lines, more effort should be put into identifying and studying physical processes that could lead to alter the He abundance with respect to that of hydrogen in giant planets. Finally, the non-detection of metastable He{\sc i} presented here may possibly be the consequence of the presence of a magnetic field, which is why it is equally important to keep attempting to directly detect exoplanetary magnetic fields.
\begin{acknowledgements}
We acknowledge financial contributions from PRIN INAF 2019 and from the agreement ASI-INAF number 2018-16-HH. IFS, MLK, MAE, MSR, IBM, AGB acknowledge the support of grant 075-15-2020-780 (GA No. 13.1902.21.0039) of the Russian Ministry of Education and Science. MLK also acknowledges the projects I2939-N27 and S11606-N16 of the Austrian Science Fund (FWF). Parallel computing simulations have been performed at Computation Center of Novosibirsk State University and SB RAS Siberian Supercomputer Center. This work has made use of data from the European Space Agency (ESA) mission {\it Gaia} (\url{https://www.cosmos.esa.int/gaia}), processed by the {\it Gaia} Data Processing and Analysis Consortium (DPAC,
\url{https://www.cosmos.esa.int/web/gaia/dpac/consortium}). Funding for the DPAC has been provided by national institutions, in particular the institutions participating in the {\it Gaia} Multilateral Agreement. We thank J. Southworth for sharing with us published data relevant to the paper. We thank the anonymous referee for their insightful comments.
\end{acknowledgements}	

\bibliographystyle{aa}
\bibliography{main}	
\begin{appendix}
\section{Tomography of the He{\sc i} transit measurements.}\label{sec:appendix_tomography}
%
\begin{figure*}[h]
\begin{center}
\includegraphics[width=17cm]{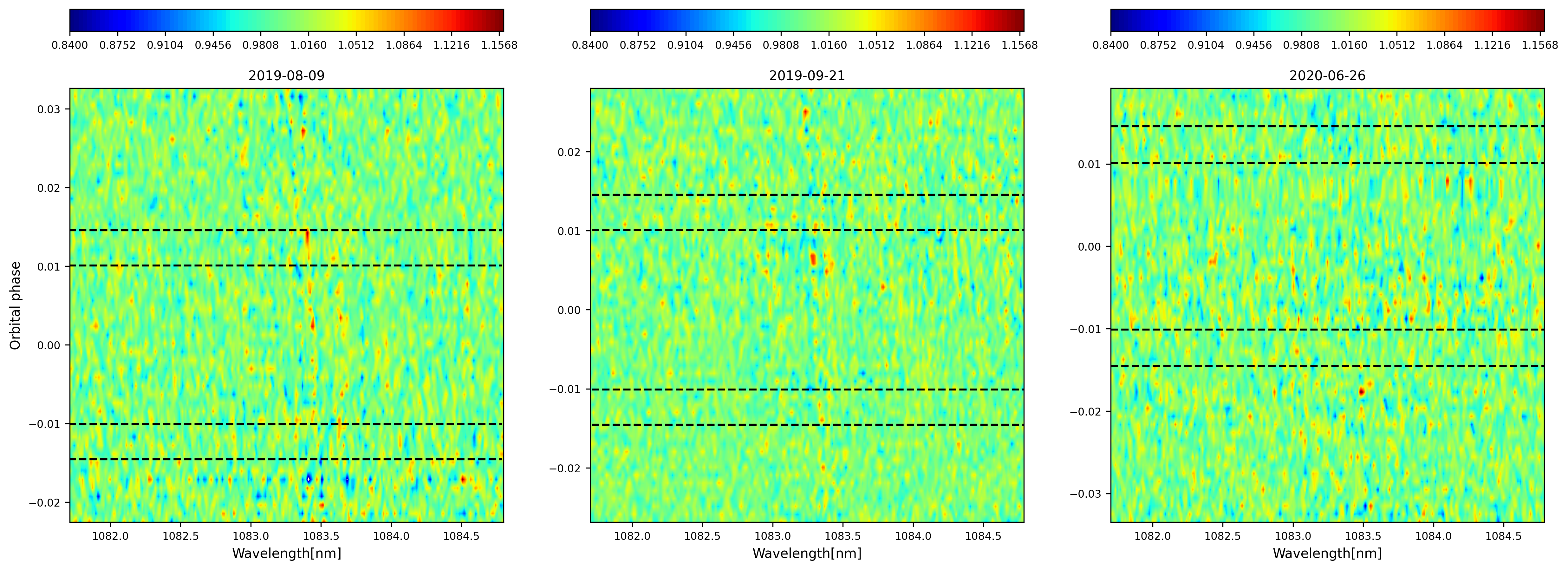}
\caption{Transmission spectra shown in tomography in the planetary rest frame for the three considered transits, as a function of wavelength and planetary orbital phase. The contact points t1, t2, t3, and t4 are marked with horizontal black lines. The feature at 1083.5~nm misaligned with the planet rest frame present in the first and second night is a residual of a telluric H$_2$O line.}
\label{fig:contour}
\end{center}
\end{figure*}
\FloatBarrier
\section{Measured HeI absorption as a function of stellar XUV irradiation and planetary surface gravity}\label{sec:appendix_HeI_summary}
%
\begin{figure}[h]
\begin{center}
\includegraphics[width=\hsize]{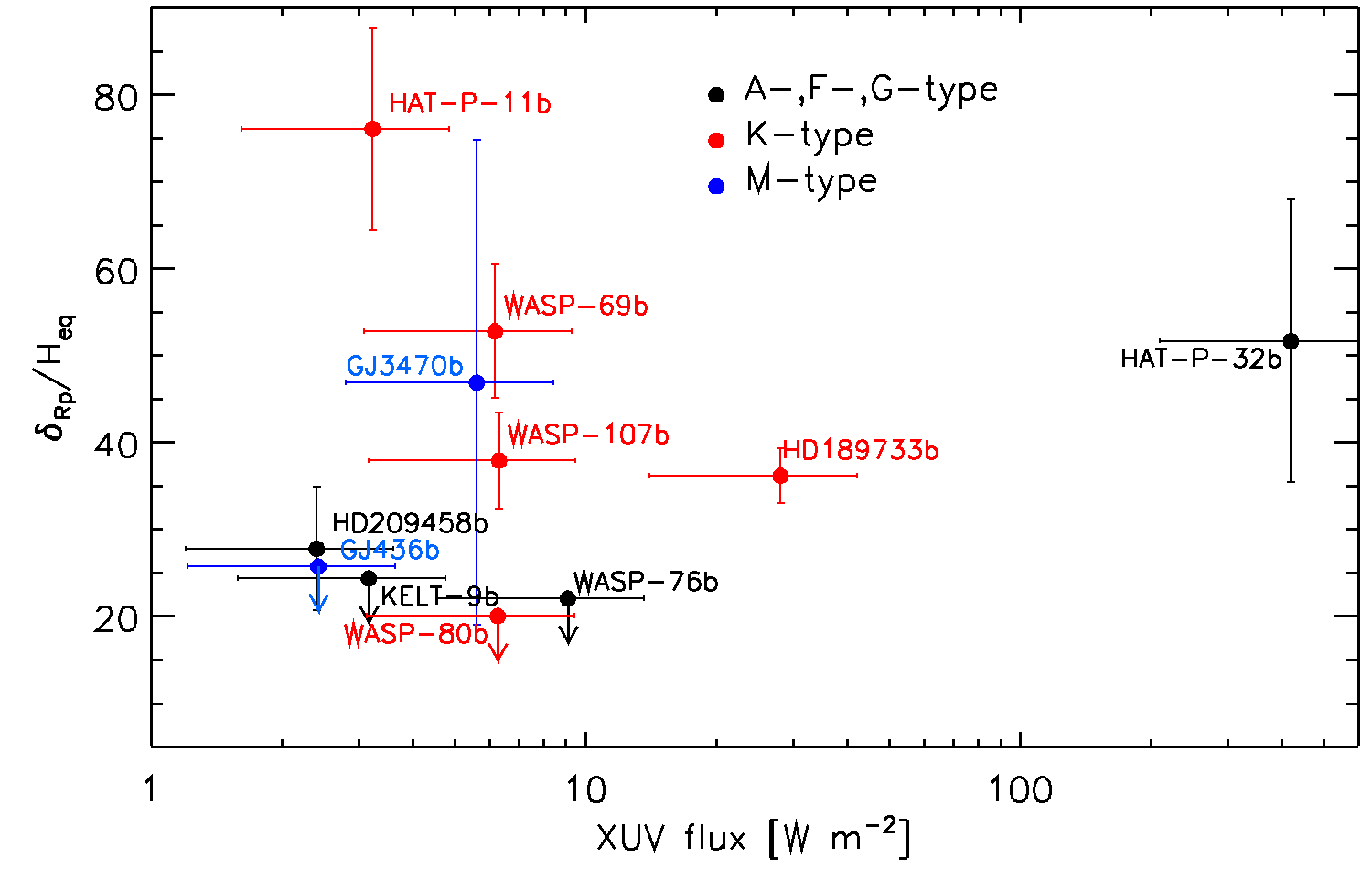}
\includegraphics[width=\hsize]{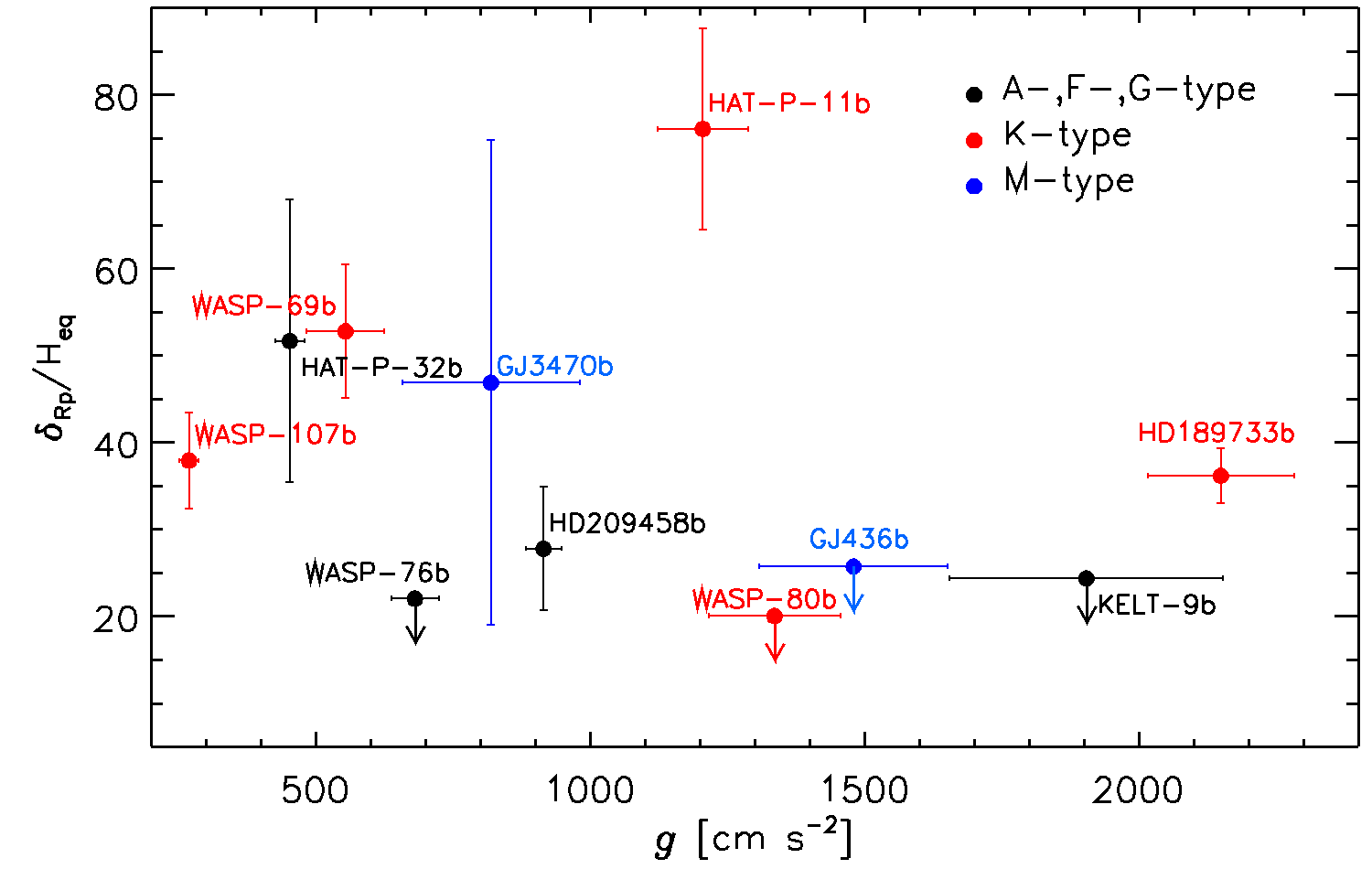}
\caption{Top: size of the measured He{\sc i} absorption signal, normalised to the atmospheric scale height computed considering the planetary parameters listed in Table~\ref{tab:systems} and a mean molecular weight of a pure hydrogen atmosphere, as a function of the incident stellar XUV flux (in logarithmic scale). Upper limits are marked by downward arrows. The symbol color indicates the spectral type of the host as given by the legend. Bottom: as the upper panel, but as a function of planetary surface gravity.}
\label{fig:He_summary}
\end{center}
\end{figure}

\end{appendix}	

\end{document}